\documentclass{article}

\usepackage[preprint,nonatbib]{neurips_2025}

\usepackage[utf8]{inputenc}
\usepackage[T1]{fontenc}

\usepackage{geometry}

\usepackage{amsmath}
\usepackage{amssymb}

\usepackage{booktabs}
\usepackage{array}
\usepackage{multirow}
\usepackage{makecell}
\usepackage{algorithm}
\usepackage{rotating} 

\usepackage{graphicx}

\usepackage{algpseudocode}

\usepackage{microtype}
\usepackage{nicefrac}

\usepackage{xcolor}
\usepackage{pifont}

\usepackage{lineno}

\usepackage{lipsum}

\usepackage{url}
\usepackage[colorlinks,linkcolor=blue,anchorcolor=blue,citecolor=blue]{hyperref}

\title{Needle-in-RAG: Prompt-Conditioned Character-Level Traceback of Poisoned Spans in Retrieved Evidence}

\author{%
  Huining Cui \\
  School of Computer Science \\
  University of Technology Sydney \\
  Sydney, Australia \\
  \texttt{Huining.Cui-1@student.uts.edu.au} \\
  \And
  Wei Liu
  \\
  School of Computer Science \\
  University of Technology Sydney \\
  Sydney, Australia \\
  \texttt{Wei.Liu@uts.edu.au} \\
}

\begin{document}

\maketitle

\begin{abstract}
Retrieval-augmented generation (RAG) improves factual grounding by conditioning large language models on retrieved evidence, but it also opens a data-layer attack surface: poisoned corpus entries can steer outputs without changing model parameters. Existing defenses and traceback methods are largely passage-level, which is too coarse for modern attacks whose effective payload may be a short fabricated claim, trigger phrase, or hidden instruction embedded inside an otherwise benign chunk. We study black-box character-level poison traceback in RAG and present RAGCharacter, a two-pass forensic framework that localizes the responsible retrieved span for a concrete misgeneration event. Pass-0 runs standard RAG while logging a prompt-anchored execution trace. Pass-1 re-enters a triggered trace and performs event-conditioned traceback over prompt-used evidence via budgeted counterfactual masking and replay, yielding an attribution span for forensic reporting and a causal span under the logged trace. We further introduce an evaluation protocol that measures both event-level chunk traceback and character-level localization fidelity. Across two QA corpora, five poisoning attack families, six target LLMs, and multiple passage- and character-level baselines, RAGCharacter achieves the best overall trade-off within our benchmark between localization accuracy and low over-attribution. These results suggest that prompt-conditioned, black-box character-level traceback can be feasible in closed-source deployment settings, moving RAG forensics from document-level suspicion toward finer-grained evidence auditing and potential remediation.
\end{abstract}

\section{Introduction}

Retrieval-augmented generation (RAG) couples a parametric language model with a retriever over a non-parametric corpus, improving knowledge-intensive generation by conditioning outputs on retrieved evidence and providing provenance \cite{Lewisretrieval,karpukhin2020dense,guu2020retrieval,izacard2021leveraging,borgeaud2022improving}. Yet even with retrieval and instruction-tuning, large language models remain prone to hallucination and brittle context use, motivating systems that more explicitly ground, verify, and attribute their answers \cite{rawte-etal-2023-troubling,huang2025survey}.

Grounding, however, expands the attack surface: corrupting the evidence pool can steer (or suppress) generation without touching model parameters, via knowledge-corruption poisoning, “blocker”/jamming documents, and indirect prompt injection embedded in retrievable text \cite{zou2025poisonedrag,zhang2025benchmarking,shafran2025machine,greshake2023not,perez2022ignore}. Existing mitigations are largely inference-time and passage-granular, e.g., isolate-then-aggregate voting, perplexity and similarity filtering, attention-based outlier removal, semantic conflict checks, query rewriting, or simply retrieving more passages, yet they degrade under adaptive adversaries and can impose substantial quality/latency costs \cite{xiang2024certifiablyrobustragretrieval,cheng2025secure,choudhary2025through,si2025secon,schwinn2023adversarial}. This suggests a complementary post-incident perspective: when a RAG system misbehaves, we should trace back the responsible poisoned evidence to explain the event and produce actionable forensic evidence, with removal or containment treated as downstream operational responses rather than the primary objective \cite{zhang2025traceback}; crucially, today’s forensics is still passage-level, leaving the within-passage poison span unidentified and forcing blunt deletion of entire chunks that may also contain benign knowledge \cite{nguyen2024enhancing}.

Rather than attempting corpus-wide poison discovery, this work studies prompt-conditioned traceback: given a concrete misgeneration event and the prompt-used evidence logged during generation, can we identify not only which retrieved passage contributed to the error, but also the implicated span within that retrieved evidence? Concretely, we define character-level localization as pinpointing the minimal contiguous text span(s) within an otherwise benign retrieved chunk, rather than flagging and removing the entire passage. This granularity is practically consequential because production RAG systems index coarse chunks for efficiency; passage-level quarantine is therefore blunt, discarding substantial clean knowledge, while leaving a short poisoned span in place may allow an exploit to remain viable under re-chunking, re-indexing, or minor query variation. Achieving character-level attribution is challenging because the poison is sparse and often deliberately camouflaged, its effect is mediated by non-linear evidence use in the generator, and retrieval introduces confounding: multiple passages may jointly support the output, while irrelevant yet similar passages dilute the signal. Finally, black-box RAG deployments often preclude gradient access and internal introspection, pushing span localization toward counterfactual, causal attribution over retrieved context \cite{koh2017understanding,pruthi2020estimating,ghorbani2019data}. While reminiscent of influence-based training-data attribution, our setting operates at inference time on retrieved evidence and should ideally remain informative under adaptive adversaries and distribution shift.

To our knowledge, current RAG-poisoning work lacks a standard character-level evaluation protocol for quantifying localization fidelity: most evaluations stop at document/passage granularity, reporting attack success rate and passage-level identification metrics without measuring whether a method actually pinpoints the responsible text inside an otherwise benign chunk \cite{zhang2025benchmarking,zhang2025traceback,zou2025poisonedrag}. This stands in contrast to explainable QA and verification benchmarks that explicitly annotate supporting sentences (evidence supervision) \cite{yang-etal-2018-hotpotqa,thorne2018fever}, and rationale benchmarks that operationalize token-/character-level overlap metrics for explanation quality \cite{deyoung2020eraser}. Existing RAG forensics and detection methods consequently provide only coarse remediation signals: traceback-style approaches return a ranked set of suspicious passages and are commonly evaluated with passage-level detection metrics \cite{zhang2025traceback}, while responsibility-attribution frameworks highlight that semantic-overlap heuristics can fail once the causal connection is paraphrased or deliberately disguised \cite{zhang2025taught}. Meanwhile, recent attacks demonstrate that poisoned content can be extremely sparse yet highly effective and stealthy (including single-document or few-shot variants) \cite{zou2025poisonedrag,chang2025one,wu2025admit}, and that adversaries can also target availability via retrieval jamming, further confounding post-hoc localization \cite{shafran2025machine}. Finally, generic explanation/attribution tools either require gradient access (often unavailable in black-box deployments) or remain too local to scale to million-document evidence pools, suggesting a gap between passage-level detection and character-level forensic attribution in RAG systems \cite{sundararajan2017axiomatic,ribeiro2016should,tan2024revprag}.

To address these challenges, we present Retrieval-Augmented Generation Cross-Span Investigation (RAGCharacter), a two-pass forensic framework that performs character-/span-level responsibility attribution for RAG. In Pass-0, RAGCharacter runs as a standard RAG pipeline but logs a complete execution trace, thereby defining an event-specific attribution scope. Once an alarm is raised, Pass-1 re-enters the recorded trace and localizes the poison within the used evidence by producing a ranked set of candidate spans. The design follows a key attribution principle articulated by black-box RAG responsibility frameworks. Accordingly, RAGCharacter combines trace-conditioned candidate pruning, answer-anchored heuristics, and replay-based influence tests to disambiguate suspicious spans within prompt-used evidence. Finally, RAGCharacter uses sanitize-and-replay counterfactuals as a validation step under the logged trace: it masks the suspected span(s) while preserving surrounding context, or drops the entire chunk as a strong baseline, and then re-queries the generator to test whether the observed misgeneration depends on the attributed evidence.

In summary, we make four contributions. (1) We introduce RAGCharacter, a black-box algorithmic framework for character-level traceback within prompt-used evidence in retrieval-augmented generation.(2) We formalize a two-pass pipeline compatible with closed-source deployment settings, consisting of Pass-0 trace logging and Pass-1 event-conditioned span attribution. (3) We introduce a character-level evaluation protocol for RAG traceback based on overlap and over-selection metrics. (4) We conduct a broad empirical evaluation across two datasets, five poisoning attack families, and six target LLM backends, focusing on traceback accuracy and localization fidelity.

By standardizing character-level evaluation and robustness-to-conditions-worsening (via curve-based reporting), our study helps narrow a gap between attack-centric benchmarks and passage-level traceback. More broadly, it reframes the RAG security question from “which document is implicated?” to “what evidence in the logged retrieved context appears responsible, and how reliably can we localize it for downstream mitigation as conditions become harder?”

\section{Related Works}
\label{sec2}

\subsection{Retrieval-Augmented Generation}

Retrieval-Augmented Generation is a paradigm that improves the factual grounding and topical coverage of LLMs by conditioning generation on evidence retrieved from an external knowledge source at inference time \cite{lewis2020retrieval, tan2024glue}. A standard RAG pipeline consists of three components: (i) a knowledge database $D$ containing textual entries (often segmented into passages/chunks), (ii) a retriever that scores and selects relevant entries from $D$ for a user query $q$, and (iii) an LLM that generates the final response conditioned on the retrieved context \cite{shafran2025machine}. In the retrieval stage, the retriever encodes the query and each candidate text into embedding vectors and ranks candidates by a similarity function, returning the top-$K$ results as external evidence \cite{tan2024revprag,shafran2025machine}. The generation stage then concatenates a system prompt, the user query, and the retrieved texts to form an augmented input, from which the LLM produces an answer \cite{zhang2025benchmarking}. 

RAG performance and behavior are heavily influenced by the retriever architecture. Prior work commonly categorizes retrieval models into bi-encoders (efficient independent encoding of queries and documents), cross-encoders (joint encoding for more accurate but expensive scoring), and poly-/multi-vector variants that trade off accuracy and efficiency via structured interaction between query and document representations \cite{cheng2024trojanrag}. This retriever--generator coupling makes RAG highly effective for factual grounding, but it also introduces new system-level considerations: because the generator trusts retrieved context, any corruption, distribution shift, or adversarial manipulation of the knowledge database can directly affect downstream generation quality and reliability \cite{zhang2025benchmarking,clop2024backdoored}. 

\subsection{RAG Poisoning Attack}
\label{subsec:rag_poisoning_attack}

RAG poisoning treats the external corpus as an attacker-controlled interface: by injecting or modifying a small number of documents, an adversary can shape what the retriever returns and thereby steer the generator \cite{chaudhari2024phantom,xue2024badrag,zhang2025traceback}. A common threat model assumes (i) write access to a portion of the knowledge source, (ii) no control over user queries, and (iii) black-box or limited access to the deployed retriever/LLM; the attacker’s objective is to maximize the probability that poisoned content appears in top-$K$ retrieval and induces a targeted output \cite{zou2025poisonedrag,zhang2025benchmarking}. Under this model, knowledge-corruption attacks formalize two necessary conditions, retrievability and generation effectiveness, and show high success with only a few injected texts \cite{zou2025poisonedrag}.

Mechanistically, attacks differ by which component they optimize. Retriever-targeting corpus poisoning crafts adversarial passages by optimizing embedding similarity, often transferring to unseen queries and even out-of-domain settings with small poisoning budgets \cite{zhong2023poisoning}. SEO-style variants sharpen the adversary goal to concept-specific query distributions and demonstrate strong visibility at extremely low poisoning rates \cite{ben2025gasliteing}. Document-level stealth attacks exploit parsing/chunking and rendering quirks to hide imperceptible payloads that survive ingestion and influence generation \cite{zhang2024human}. System-level attacks expand objectives beyond misinformation: trigger-based backdoors create persistent trigger--behavior shortcuts \cite{cheng2024trojanrag}, retriever backdoors poison fine-tuning to force attacker-chosen instruction retrieval \cite{clop2024backdoored}, and DoS/jamming shows a single blocker document can reliably elicit refusals for targeted queries \cite{shafran2025machine}. Recent benchmarks further argue that single-attacker ASR can be misleading in realistic settings with multiple competing poisoners, motivating competitiveness-based evaluation \cite{chen2025poisonarenauncoveringcompetingpoisoning,zhang2025benchmarking}. Finally, decoding-time optimization can generate fluent poisoned documents that satisfy both retrievability and generation control, weakening defenses that rely on detecting low-quality or high-perplexity text \cite{zhang2024adversarial}.

\subsection{Traceback of Poisoned Texts in RAG}
\label{sec:traceback}

RAG poisoning is increasingly realized through data-layer manipulations that implant malicious or misleading content into the external knowledge base, thereby steering retrieval and downstream generation \cite{zhong2023poisoning}. Beyond corpus-level corruption, attacks can also target the retrieval pipeline itself\cite{clop2024backdoored}, or exploit long-lived knowledge stores with persistent and query-triggered behaviors \cite{zou2025poisonedrag,cheng2024trojanrag}. Importantly, modern poisoning can be deliberately stealthy: it may take the form of plausible typos that change model behavior while remaining human-natural \cite{cho2024typos}, or even ``human-imperceptible'' payloads that hide instructions in visually invisible text \cite{zhang2024human}. These diverse threat surfaces motivate not only online mitigation but also post-incident traceback that identifies and cleans the poisoned sources. 

Most existing defenses emphasize inference-time mitigation: they attempt to suppress the impact of poisoned contexts through filtering, rewriting, or robustness-oriented aggregation \cite{zhang2025benchmarking,si2025seconrag}. Benchmark-style studies further suggest that these defenses can be brittle under realistic poisoning settings, motivating an operational need for systematic root-cause analysis and durable data remediation beyond one-off mitigation \cite{zhang2025benchmarking,zhang2025traceback}.

A major family of defenses applies coarse-grained passage or chunk-level filtering, using perplexity- or embedding-based scores to flag suspicious retrieved texts \cite{zhang2025benchmarking,zhang2025traceback}. Process-optimized defenses modify the RAG workflow, e.g., paraphrasing, reranking, or heuristic prompt controls, to reduce the effective attack surface, but still treat retrieved items as largely atomic evidence units rather than localizing the specific causal content within them \cite{zhang2025benchmarking}. Robust aggregation methods aim to tolerate corruption by isolating evidence and aggregating across retrieved candidates, often relying on voting/consistency assumptions that may degrade when poisoning is strong, dense, or strategically crafted \cite{shen2025reliabilityrag,si2025seconrag}. Collectively, these mitigation-oriented strategies can reduce immediate harm, yet they often do not directly answer the forensic question of what exactly should be removed or repaired in the resident knowledge store after an incident \cite{zhang2025traceback}.

Complementary work targets indirect prompt injection by detecting and excising malicious instructions from retrieved external content \cite{Wen_2025,li-etal-2025-piguard}. Instruct-Detector leverages model-derived signals to identify hidden instructions and discard suspicious documents before they are passed to the LLM \cite{Wen_2025}, and further sketches a finer-grained option via recursive segmentation to remove injected portions while preserving benign context \cite{Wen_2025}. Prompt-guard models investigate over-defense driven by trigger-word shortcuts and introduce token-level rechecking to reduce false positives \cite{li-etal-2025-piguard}. Despite these advances, the primary objective remains preventive screening: these methods aim to block or sanitize malicious inputs at inference time, but typically do not produce an explicit responsibility map that supports post-incident traceback and long-term corpus remediation \cite{zhang2025traceback,si2025seconrag}.

To clarify the landscape, Table~\ref{tab:literature_rev} organizes representative prior work along two dimensions: research scope and poison attribution level. The table highlights that most existing traceback and removal mechanisms operate at coarse granularity, while only a minority of approaches reason below the passage level, and rarely as the primary forensic artifact.

In summary, the literature exposes a practical granularity gap: remediation actions are typically performed at coarse units, even though modern poisoning can be triggered by subtle, localized manipulations that do not align cleanly with document- or chunk-level boundaries. This mismatch limits the precision and minimality of post-hoc detoxification: operators are forced to choose between over-removal (hurting utility) and under-removal (leaving residual attack surface). These observations motivate the need for finer-grained forensic traceback in RAG, potentially supporting lower-collateral-damage remediation in deployment-relevant settings.

\begin{table}[hbpt]
\centering
\normalsize
\caption{Research scope and attribution level comparison of previous works. (RAGCharacter performs post-hoc traceback on triggered events and supports downstream remediation analysis, but is not itself an online defense or standalone removal mechanism.)}
\label{tab:literature_rev}
\renewcommand{\arraystretch}{1.2}
\resizebox{\textwidth}{!}{%
\begin{tabular}{@{}l@{}cccccccccc@{}}
\toprule
\multirow{3}{*}{\textbf{Name}}
  & \multicolumn{3}{c}{\textbf{Research Scope}}
  & \multicolumn{6}{c}{\textbf{Attribution Level}} \\
\cmidrule(lr){2-4} \cmidrule(lr){5-11}
& \makecell{RAG Poisoning} 
& \makecell{Poison \\Detection} 
& \makecell{Defence or \\ Poison Removal} 
& \makecell{Document/ Passage} 
& \makecell{Entry/ Chunk} 
& \makecell{Segment} 
& \makecell{Keyword} 
& \makecell{Token} 
& \makecell{Character} \\
\midrule
Machine Against the RAG\cite{shafran2025machine} & \checkmark & \checkmark & \checkmark & \checkmark & & & & &  \\
RevPRAG\cite{tan2024revprag} &  &\checkmark&  &  & &  &  &\checkmark&\\
PoisonedRAG \cite{zou2025poisonedrag} & &\checkmark &\checkmark & &\checkmark & &  &  &  &  \\
Agent Poison \cite{10.5555/3737916.3742052} &\checkmark &\checkmark &\checkmark & \checkmark &  &  &  &  &  &  \\
NeuroGen Poisoning\cite{zhuneurogenpoisoning} &\checkmark &  &  &  & & &  &  & \\
PR-Attack \cite{10.1145/3726302.3730058} &\checkmark &  &  &  & &  &  &  & \\
RAG Forensics \cite{zhang2025traceback} & &  & \checkmark& & \checkmark&  &  &  & \\
Reliability RAG\cite{shen2025reliabilityrag} & & \checkmark & \checkmark & \checkmark &  &  &  &  &  \\
SeCon-RAG\cite{si2025seconrag} & & \checkmark & \checkmark & \checkmark &  &  &  &  &  \\
One Shot Dominance \cite{chang-etal-2025-one} & \checkmark &  &  &  &  & &  &  &  \\
The RAG Paradox\cite{choi2025ragparadoxblackboxattack}& \checkmark &  &  &  &  & &  &  &  \\

PoisonedEye\cite{zhang2025poisonedeye} & \checkmark &  &  &  &  & &  &  &  \\

Scalable Data Extraction \cite{qi2024follow} & \checkmark &  &  &  &  & &  &  &  \\

AIP\cite{chaturvedi-etal-2025-aip} && \checkmark &  &  &  &  & &  &  &  \\

Instruct Detector\cite{Wen_2025} &  &\checkmark&\checkmark&  &\checkmark&  &  &  &  &  \\

RAG-SAGE\cite{zeng-etal-2025-mitigating} & \checkmark &\checkmark&\checkmark&  &\checkmark& &  &  &  &  \\

RAG Privacy\cite{zeng-etal-2024-good} & \checkmark &  &\checkmark&  &  & \checkmark &  &  &  \\

Context Influence\cite{flemings-etal-2025-estimating} & &  &\checkmark &  &  &  &  &\checkmark& \\

PIA-Defense\cite{chen-etal-2025-defense} & & \checkmark & &\checkmark &  & &  & &  \\

PIGuard\cite{li-etal-2025-piguard} & & \checkmark &  &  &  & \checkmark &  &  &  \\

BIPIA\cite{10.1145/3690624.3709179} & \checkmark &  &  &  &  &  &  &  & \\

CorpusPoison\cite{zhong2023poisoning} & &\checkmark &\checkmark &\checkmark &  & &  &  & \\

ContentPoison\cite{zhang2024human} &  &  &\checkmark &\checkmark &  & &  &  &  \\

AbvDecoding\cite{zhang2024adversarial} & &  &  & \checkmark &  &  &  &   \\

GASLITE\cite{ben2025gasliteing} &  &\checkmark &\checkmark &\checkmark &  & &  &  &   \\

GARAG\cite{cho2024typos} & \checkmark &\checkmark &  &\checkmark&  & &  &  &  &  \\

PoisonArena\cite{chen2025poisonarenauncoveringcompetingpoisoning} & \checkmark &\checkmark &\checkmark &  &\checkmark & &  &  &  \\

TrojanRAG\cite{cheng2024trojanrag} & \checkmark &\checkmark &  &\checkmark &  & &  &  &  &  \\

RAG Security Bench\cite{zhang2025benchmarking} & \checkmark &\checkmark &\checkmark &  & \checkmark & &  &  &   \\

BRPI-RAG\cite{clop2024backdoored} & \checkmark &  &  &  &  & &  &  &   \\

PoisonRAG-ROBUST\cite{su2025robustretrievalaugmentedgenerationevaluating} & \checkmark &  &  &  &  & &  &  &  \\

PANDORA\cite{deng2024pandorajailbreakgptsretrieval} & \checkmark &  &  &  &  &  &  &  &   \\
\midrule
\textbf{RAGCharacter (Present Work)} &  & \checkmark &  &  & \checkmark & \checkmark &  &  & \checkmark \\
\bottomrule
\end{tabular}%
}
\end{table}

\section{Character-level Poison Traceback Problem Statement}
\label{subsec:traceback}

\subsection{Example of Character-level Traceback in RAG}

\begin{figure}[t!]
\centering
\includegraphics[width=\textwidth]{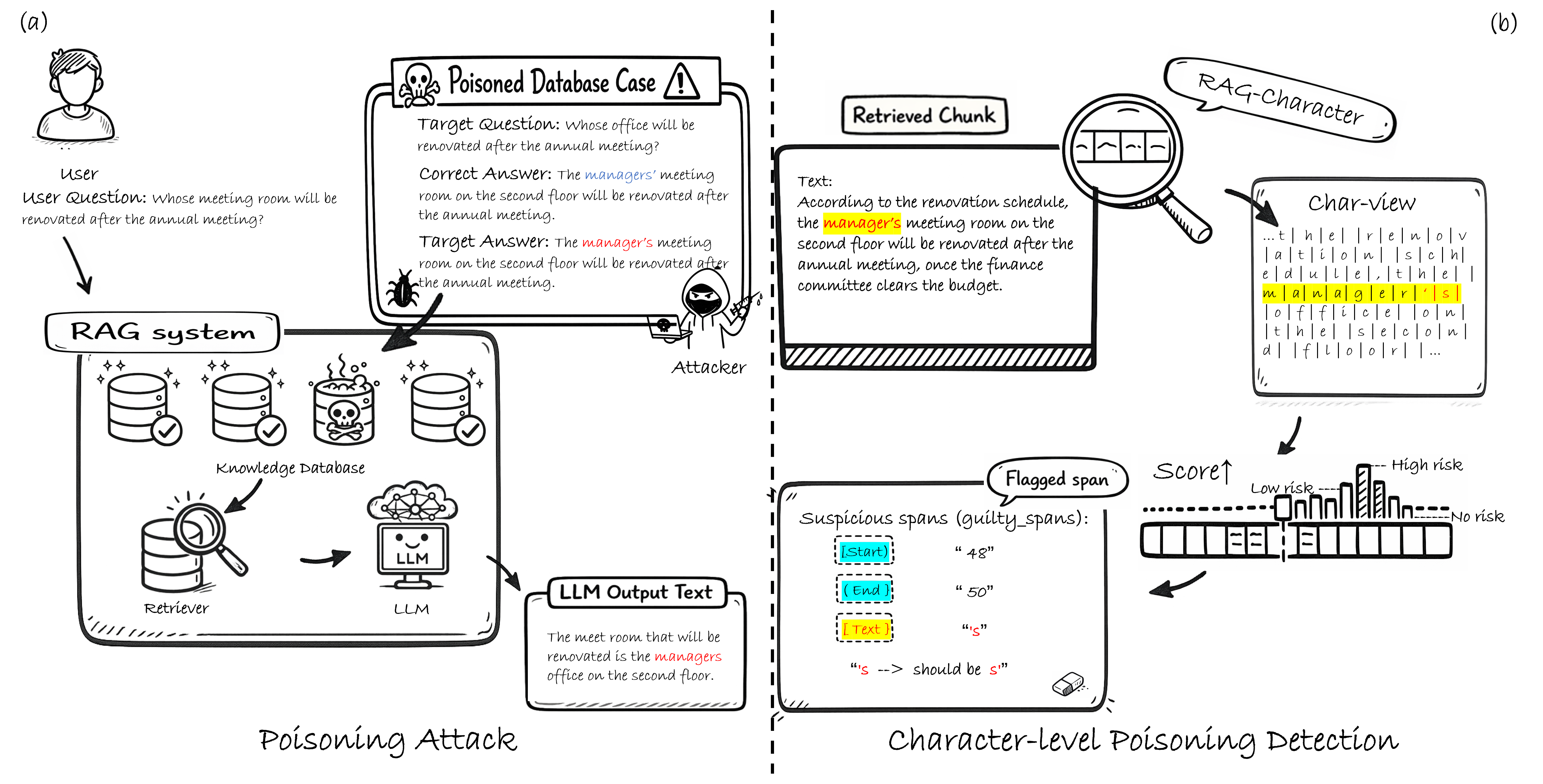}
\caption{RAG Poisoning Attack and Character-level Detection.}
\label{End-to-End Illustration}
\end{figure}

Document-level and entry-level scoring is often too coarse for RAG forensics: poisoning artifacts may be localized to short manipulative phrases that occupy a small fraction of a retrieved context (Figure~\ref{End-to-End Illustration} (a)). Treating each retrieved document as a monolithic unit amplifies benign content and obscures the causal evidence that drives the poisoned answer. We therefore adopt a character-level view (Figure~\ref{End-to-End Illustration} (b)), tracing suspicious text fragments within retrieved contexts and using them as forensic evidence for attribution.

\subsection{Threat Model}
\label{subsec:threat_model}

\paragraph{System setting.}
We consider a deployed RAG service consisting of a knowledge database $D$, a retriever $\mathcal{R}$, and an LLM $\mathcal{G}$. Given a user query $q$, the retriever returns a ranked top-$K$ set of hits $H=\{(c_j,s_j)\}_{j=1}^{K} \leftarrow \mathcal{R}(q;D)$, and the generator produces an output $y \leftarrow \mathcal{G}(q,H_U)$, where $H_U \subseteq H$ denotes the subset of retrieved hits actually inserted into the prompt. The service provider (operator) maintains $D$ and its indexing pipeline, and can update, remove, or replace content in $D$ as part of remediation \cite{zhang2025traceback}.

\paragraph{Attacker.}
Following prior RAG poisoning literature, we assume an attacker can poison the knowledge database by injecting or modifying a small number of texts such that the RAG system returns attacker-desired behaviors for targeted queries \cite{zou2025poisonedrag,zhang2025benchmarking,cheng2024trojanrag}. We model the attacker goal as inducing outputs that satisfy an attacker-specified predicate when poisoned evidence is retrieved and consumed by the LLM \cite{zou2025poisonedrag,shafran2025machine,cheng2024trojanrag}. We consider both white-box and black-box poisoning settings commonly used in prior work, where the attacker may or may not have access to the retriever architecture/parameters used to craft highly retrievable poisons, but does not require interaction with the deployed service \cite{zou2025poisonedrag,zhang2025benchmarking}. The poisoned payload may be overt (misinformation) or stealthy/obfuscated, including surface-form manipulations such as plausible typos or visually imperceptible injected content that survives ingestion and chunking \cite{cho2024typos,zhang2024human}.

\paragraph{Traceback operator.}
We assume the traceback system is deployed by the service provider that operates the RAG pipeline. Accordingly, it has (i) full read/write access to $D$ and the ability to rebuild the retrieval index, and (ii) query access to the production retriever and LLM in closed-source deployment settings. Crucially, to reflect common real-world deployments, we assume the retriever and LLM may be closed-source and thus the traceback system does not require access to internal parameters, gradients, or training data. 
We also assume no prior knowledge of the attacker’s template, trigger, or poisoning strategy.

\paragraph{Misgeneration events and logging.}
In this work, a misgeneration denotes any observed RAG output that is flagged as problematic with respect to the intended behavior of the system. This includes incorrect factual answers, attacker-induced target answers, unsupported refusals, or other outputs that are inconsistent with the trusted use of retrieved evidence. Importantly, the term is operational rather than oracle-based: a misgeneration is defined by the occurrence of a suspicious event \((q,y)\), not by prior access to a gold answer. The traceback stage then determines whether the event is attributable to poisoned prompt-used evidence or is better explained by model-internal error.

The operator collects misgeneration events via user feedback or monitoring. Each event contains the query $q$ and the produced output $y$ flagged as problematic, together with the retrieval trace needed to reproduce the run, such as document identifiers, chunk boundaries, and the exact evidence strings used in the prompt. We assume users report events honestly, and we focus on determining whether such events are attributable to poisoning rather than purely model-internal errors.

\paragraph{Out of scope.}
We do not model attacks that directly compromise the internal parameters or execution environment of the retriever/LLM as the primary root cause; our focus is on evidence-side poisoning stored in $D$. We also exclude infrastructure-level tampering and adversarial feedback generation.

\subsection{Key Challenges}
\label{subsec:key_challenges}
Character-level poison traceback is fundamentally more challenging than passage- or chunk-level forensics because the causal signal is both fine-grained and system-mediated. In a RAG pipeline, influence propagates through multiple discrete stages, tokenization, retrieval, and generation, making it difficult to isolate which parts of the stored evidence are truly responsible for a misgeneration event. Under the threat model described in \S\ref{subsec:threat_model}, we identify three key challenges that any practical character-level traceback system must address.


\paragraph{(C1) Black-box and non-differentiable components.} Unlike training-phase data poisoning, RAG poisoning does not directly alter model parameters. Responsibility arises only through discrete retrieval and generation behaviors, often in closed-source deployment settings. This rules out gradient-based or parameter-level attribution methods commonly used in classical poisoning forensics.

\paragraph{(C2) Retrieval confounding effects.} Character-level sanitization can change token boundaries, embeddings, and retrieval rankings. As a result, a remediation action may suppress a misgeneration because it alters retrieval behavior rather than because it removes the true causal substring, complicating counterfactual reasoning about responsibility.

\paragraph{(C3) Minimality and ambiguity of remediation.} Not all incorrect outputs are caused by poisoning; some stem from intrinsic model limitations or incomplete knowledge. Coarse-grained removal risks unnecessary loss of benign information, while under-removal leaves residual attack surface. Character-level traceback must therefore support minimally invasive remediation under uncertainty.

\section{RAGCharacter for Character-level Poison Traceback}

\subsection{RAGCharacter Architecture}
\label{sec:ragcharacter-arch}

RAGCharacter is a two-pass traceback system that localizes poisoned evidence at character granularity while remaining compatible with black-box LLM deployments, with core actions shown in Algorithms~\ref{alg:pass0} and ~\ref{alg:pass1}. We adopt the forensic setting of prior traceback works, where the defender is not assumed to know the attacker’s injection strategy and poisoned entries may be arbitrarily distributed throughout the knowledge store. Given a user query \(q\) and an observed misgeneration event, RAGCharacter identifies responsible retrieved content and attempts to localize compact character spans that best explain the event under the logged trace, yielding an auditable forensic artifact; masking or removal is treated as a downstream operational option.

\paragraph{Algorithm.}
We assume a standard retrieval-augmented generation pipeline consisting of a retriever \(\mathcal{R}\) over a corpus \(\mathcal{D}\) and a generator \(\mathcal{G}\). For each query \(q\), the retriever returns a ranked list of hits \(H=\{(c_j,s_j)\}_{j=1}^{K}\), where \(c_j\) is a chunk and \(s_j\) is its retrieval score. The generator consumes a truncated subset \(H_U=\textsc{TopU}(H)\) to produce an output \(y=\mathcal{G}(q,H_U)\). In contrast to mitigation-centric defenses, RAGCharacter is designed for forensics-first traceback: it treats the observed generation event \((q,y)\) as the primary evidence and aims to recover a responsibility chain
\[
(q,y)\ \rightarrow\ (c^\star)\ \rightarrow\ (\ell^\star,r^\star),
\]
where \(c^\star\) is the top-ranked implicated chunk among \(H_U\), and \((\ell^\star,r^\star)\) is a character-level span inside \(c^\star\).

\begin{algorithm}[hbpt]
\caption{\textsc{RAGCharacter} Pass-0: RAG Run and Prompt-Anchored Trace Logging}
\label{alg:pass0}
\small
\begin{algorithmic}[1]
\Statex \textbf{Input:} Corpus $\mathcal{D}$; queries $Q$; retriever; generator; top-$K$.
\Statex \textbf{Output:} Full trace set $\mathcal{T}$ and triggered trace set $\mathcal{T}^{+}$ for Pass-1.

\For{each query $q$ in $Q$}
    \State Find the top-$K$ ranked hits $H$ for $q$ from the corpus $\mathcal{D}$.
    \State Generate an answer $y$ from $q$ and $H$.
    \State Create a trace $\tau$ containing $q$, $H$, $y$, and their character offsets.
    \State Add $\tau$ to the full trace set $\mathcal{T}$.
    \If{$y$ overlaps with the text of some prompt-used chunk in $H$}
        \State Add $\tau$ to the triggered trace set $\mathcal{T}^{+}$ for Pass-1.
    \EndIf
\EndFor
\State \Return $(\mathcal{T},\mathcal{T}^{+})$.
\end{algorithmic}
\end{algorithm}

\begin{algorithm}[hbpt]
\caption{\textsc{RAGCharacter} Pass-1: Event-Conditioned Character-Level Traceback}
\label{alg:pass1}
\small
\begin{algorithmic}[1]
\Statex \textbf{Input:} 
A triggered trace $\tau\in\mathcal{T}^{+}$; query $q\in\tau$; top-K ranked hits $H\in\tau$; generator $\mathcal{G}$; masking function $\textsc{Mask}$; cumulative masked regions $M$ initialized as $\emptyset$; chunk budget $B_c$; sentence budget $B_s$; focus-region budget $B_t$; span budget $B_p$; bisection budget $T_b$; minimum span length $\eta$.
\Statex \textbf{Output:} Updated trace $\tau'$.

\For{each prompt-used chunk $h_i$ among the top $B_c$ candidates} \Comment{Step 1: Chunk screening}
    \State Mask $h_i$ in the prompt and replay the model: $\hat{y}_{-h_i}^{(t)} \leftarrow \mathcal{G}(q,\textsc{Mask}(H,h_i))$.
    \State Record how much masking $h_i$ weakens the event.
\EndFor
\State Let $c$ be the chunk whose removal weakens the event the most. \Comment{Step 2: Sentence screening}

\For{each candidate sentence $s_i$ among the top $B_s$ sentences in $c$}
    \State Mask $s_i$ and replay the model: $\hat{y}_{-s_i}^{(t)} \leftarrow \mathcal{G}(q,\textsc{Mask}(H,s_i))$.
    \State Record how much masking $s_i$ weakens the event.
\EndFor
\State Keep up to $B_t$ best sentence regions. \Comment{Step 3: Phrase screening}

\For{each finer phrase span $p_i$ among the top $B_p$ candidates in the retained sentence regions}
    \State Mask $p_i$ and replay the model: $\hat{y}_{-p_i}^{(t)} \leftarrow \mathcal{G}(q,\textsc{Mask}(H,p_i))$.
    \State Record how much masking $p_i$ weakens the event.
\EndFor
\State Let $p$ be the phrase span that weakens the event the most. \Comment{Step 4: Character screening}

\For{$j=1$ to $T_b$}
    \State Bisect $p$ and keep the shorter subspan that better preserves event suppression.
    \If{the length of $p$ is at most $\eta$}
        \State \textbf{break}
    \EndIf
\EndFor
\State Record the resulting span $p$ as $\pi_{\mathrm{causal}}$ and add it to $M$. 

\State Replay the model on the final masked prompt and record the sanitized output as $y'\leftarrow \mathcal{G}(q,\textsc{Mask}(H, M))$.
\State Write $c$, $\pi_{\mathrm{causal}}$, $M$, and $y'$ back into the trace to obtain $\tau'$.
\State \Return $\tau'$.
\end{algorithmic}
\end{algorithm}

\subsection{Pass-0: event logging with prompt-anchored evidence}
Pass-0 corresponds to a normal RAG run augmented with deterministic tracing. For each query, RAGCharacter logs: (i) the top-\(K\) retrieval list with identifiers, (ii) the exact subset \(H_U\) inserted into the prompt, (iii) the fully rendered prompt and the model completion \(y\), and (iv) prompt offsets that map each retrieved chunk back to prompt character positions. This prompt anchoring enables later attribution spans to be expressed both in chunk coordinates \((\ell,r)\) and prompt coordinates \((\ell_p,r_p)\), making results reproducible and auditable.

A key design choice is to define a traceback event independently of ``ground-truth correctness.'' In realistic poisoned settings, the system may lack verified answers and, under multi-poison competition, masking one poison does not necessarily restore correctness. RAGCharacter therefore defines events in terms of adoption of retrieved poison candidates: an event occurs when the generated answer aligns with an incorrect answer string supported by some retrieved chunk in \(H_U\). This event-level framing aligns with forensic objectives and avoids conflating traceback quality with end-task accuracy.

\subsection{Pass-1: Event-Conditioned Character-Level Traceback}
\label{sec:pass1_detail}

Pass-1 takes as input a single Pass-0 trace rather than the entire corpus. Let \(H_U=\{b_j\}_{j=1}^{U}\) denote the prompt-used hits that were actually inserted into the generation prompt, where each prompt block \(b_j\) stores a chunk identifier, retrieval rank, and prompt-global character offsets \([a_j,e_j)\). If the trace does not explicitly record \(U\), our implementation falls back to \(U=10\). Hence, Pass-1 never searches the full database; it only re-enters the small set of evidence blocks that were truly exposed to the generator in Pass-0.

Let \(P^{(0)}\) denote the rendered prompt-side evidence string and let \(z^{(0)}\) be the baseline event signals logged in Pass-0, including the model prediction \(\hat{y}^{(0)}\) and event indicators such as \texttt{PAR}, \texttt{T\_ASR}, and \texttt{CONFUSION}. Pass-1 is invoked only when a user-chosen trigger predicate \(\mathsf{Trig}(z^{(0)})\) is satisfied; by default we use \texttt{PAR}, although the implementation also supports \texttt{T\_ASR}, \texttt{CONFUSION}, \texttt{ANY}, and \texttt{ALWAYS}. Importantly, if these signals are already present in the Pass-0 trace, Pass-1 reuses them directly and avoids an extra baseline LLM call.

At round \(t\), Pass-1 maintains a set of previously selected mask regions \(M^{(<t)}\) and constructs the current replay prompt
\[
P^{(t)}=\textsc{Mask}\!\left(P^{(0)}, M^{(<t)}\right).
\]
The current prediction under \(P^{(t)}\), denoted \(\hat{y}^{(t)}\), becomes the event anchor for all subsequent candidate pruning in that round. By default, we allow at most three rounds, which was sufficient for most single-span and small multi-span cases in our experiments.

\paragraph{Answer-anchored fast path.}
Before running broader scans, Pass-1 attempts a cheap one-step traceback. It finds the highest-ranked prompt block whose text contains \(\hat{y}^{(t)}\) under normalized matching (case/whitespace-robust containment), and extracts a deterministic attribution span \(\pi_{\text{attr}}^{(t)}\) inside that block using the following priority rule: (i) exact case-insensitive match of \(\hat{y}^{(t)}\); otherwise (ii) the leading capitalized phrase of 1--6 words; otherwise (iii) the first non-space token. We then mask only this span and replay once. If the trigger is resolved, Pass-1 terminates the round immediately. In the implementation, this shortcut corresponds to the \texttt{attribution\_first} branch and often reduces Pass-1 to a single additional model call.

\paragraph{Chunk-level occlusion screening.}
If the fast path fails, Pass-1 performs counterfactual chunk screening. To save cost, we first restrict testing to blocks that already contain the current bad answer:
\[
\mathcal{B}^{(t)}=
\begin{cases}
\{b\in H_U:\textsc{ContainsNorm}(b,\hat{y}^{(t)})\}, & \text{if non-empty},\\
H_U, & \text{otherwise},
\end{cases}
\qquad
|\mathcal{B}^{(t)}|\leq 5,
\]
where the upper bound follows from the default setting \texttt{max\_chunks\_to\_test}=5. For each \(b\in\mathcal{B}^{(t)}\), we mask its entire prompt region \([a_b,e_b)\), re-query the generator, and compute an influence score
\[
s_{\text{chunk}}(b)=\textsc{Influence}\!\left(z^{(t)}, z_{b}^{(t)}; \texttt{objective}\right),
\]
where \(z^{(t)}\) and \(z_b^{(t)}\) are the pre-mask and post-mask signals, respectively. The default objective is \texttt{event}, which is label-free and scores candidates only by how strongly they suppress the observed event; an \texttt{oracle} mode is retained for analysis when gold/target labels are available. A candidate is marked successful when the masked replay no longer satisfies the trigger predicate. Chunks are ranked by \(s_{\text{chunk}}(b)\), with retrieval rank used only as a tie-breaker. By default, we enable early stopping and stop the scan as soon as a successful mask is found.

\paragraph{Sentence-first refinement.}
Let \(c^\star\) denote the highest-scoring chunk selected above. Inside \(c^\star\), we next run a sentence-level search before descending to arbitrary character spans. Concretely, we split the raw chunk text into at most 50 sentence spans, drop sentences shorter than 8 characters, ignore candidates overlapping earlier masks, and sort the remaining sentences by three keys: whether they contain \(\hat{y}^{(t)}\), sentence length (longer first), and position (earlier first). Only the top 5 sentences are replay-tested by one-sentence masking. If a sentence mask succeeds, we do not immediately accept the whole sentence; instead, we send that sentence to a bisection routine that shrinks it to a shorter causal span. Even when no sentence directly succeeds, sentence scoring is still useful: we retain the top 2 sentence regions and, whenever possible, restrict later fine-grained span search to those regions only.

\paragraph{Fine-grained span generation and scoring.}
Hard cases are handled by an event-conditioned character-span proposal stage. Given the selected chunk \(c^\star\), the routine \textsc{CandidateSpansForChunk} generates a small candidate pool
\[
\mathcal{C}^{(t)}=\textsc{CandidateSpansForChunk}\!\left(c^\star,\hat{y}^{(t)}\right),
\qquad
|\mathcal{C}^{(t)}|\leq 12,
\]
where the bound comes from \texttt{max\_span\_candidates}=12. Candidates overlapping previously masked regions are discarded. If sentence-level scores are available, we further filter \(\mathcal{C}^{(t)}\) to spans lying inside the retained top-2 sentence regions whenever this restriction leaves at least one valid candidate. Each span \(u\in\mathcal{C}^{(t)}\) is then evaluated by masked replay and assigned an influence score
\[
s_{\text{span}}(u)=\textsc{Influence}\!\left(z^{(t)}, z_{u}^{(t)}; \texttt{event}\right).
\]
The best span becomes the seed for final refinement. If no valid fine span can be proposed, Pass-1 falls back to masking the entire selected chunk, which ensures that the sanitize-and-replay loop remains well-defined even in degenerate cases.

\paragraph{Bisection to a minimal causal span.}
The implementation distinguishes two outputs. The first is the deterministic attribution span \(\pi_{\text{attr}}\), used for reporting and character-level overlap evaluation. The second is the causal span \(\pi_{\text{causal}}\), chosen by masked replay and then minimized by \textsc{BisectMinimalSpan}. Starting from either a successful sentence mask or the best fine-grained proposal, the bisection stage performs at most 6 refinement steps and never returns a span shorter than 4 characters (\texttt{max\_bisect\_steps}=6, \texttt{min\_span\_len}=4). The final span is stored in both chunk-local coordinates and prompt-global coordinates, together with its verbatim snippet, retrieval rank, and post-mask event signals. This dual-coordinate logging makes the output directly auditable and easy to map back to the indexed chunk.

\paragraph{Iterative sanitize-and-replay.}
After selecting \(\pi_{\text{causal}}^{(t)}\), Pass-1 appends it to the cumulative mask set \(M\), sanitizes the prompt again, and re-evaluates the model. If the trigger is still active, the newly produced prediction becomes \(\hat{y}^{(t+1)}\), and the next round starts from the already-masked prompt. Thus, Pass-1 can return either a single span (\texttt{mask\_span}) or a small set of spans (\texttt{mask\_multi}) when the poisoned behavior depends on multiple local cues. By default, the loop stops as soon as the trigger becomes false or the round budget \(R=3\) is exhausted.

\paragraph{Complexity and practical budget.}
The search budget is tightly bounded: reuse Pass-0 signals when available, test at most 5 chunks, at most 5 sentences, at most 12 fine-grained span proposals, and cache repeated masked-prompt evaluations. As a result, Pass-1 remains compatible with closed-source deployment settings while still producing a detailed forensic artifact containing all chunk tests, sentence tests, span tests, selected masks, and the final sanitized replay.

\paragraph{Implementation properties.}
RAGCharacter is compatible with settings where only query access to the generator is available. Pass-0 requires no additional model calls beyond normal RAG generation. Pass-1 is triggered and budgeted: it runs only on traceable events and restricts the search space to the prompt-used retrieval set \(H_U\), keeping search cost bounded by the configured budgets. By recording prompt offsets and chunk identifiers, RAGCharacter produces a reproducible trace artifact for triggered events: under the logged prompt, the system returns a chunk and character interval that best explain the observed event. In summary, RAGCharacter extends prior document- and chunk-level traceback systems by introducing a span-aware forensic architecture by introducing a span-aware forensic architecture that (i) logs prompt-anchored evidence at generation time and (ii) localizes responsibility at character granularity at analysis time; answer recovery is not assumed and is treated as a separate, secondary question in multi-poison and no-ground-truth regimes.

\begin{figure}[hbpt]
\centering
\includegraphics[width=\textwidth]{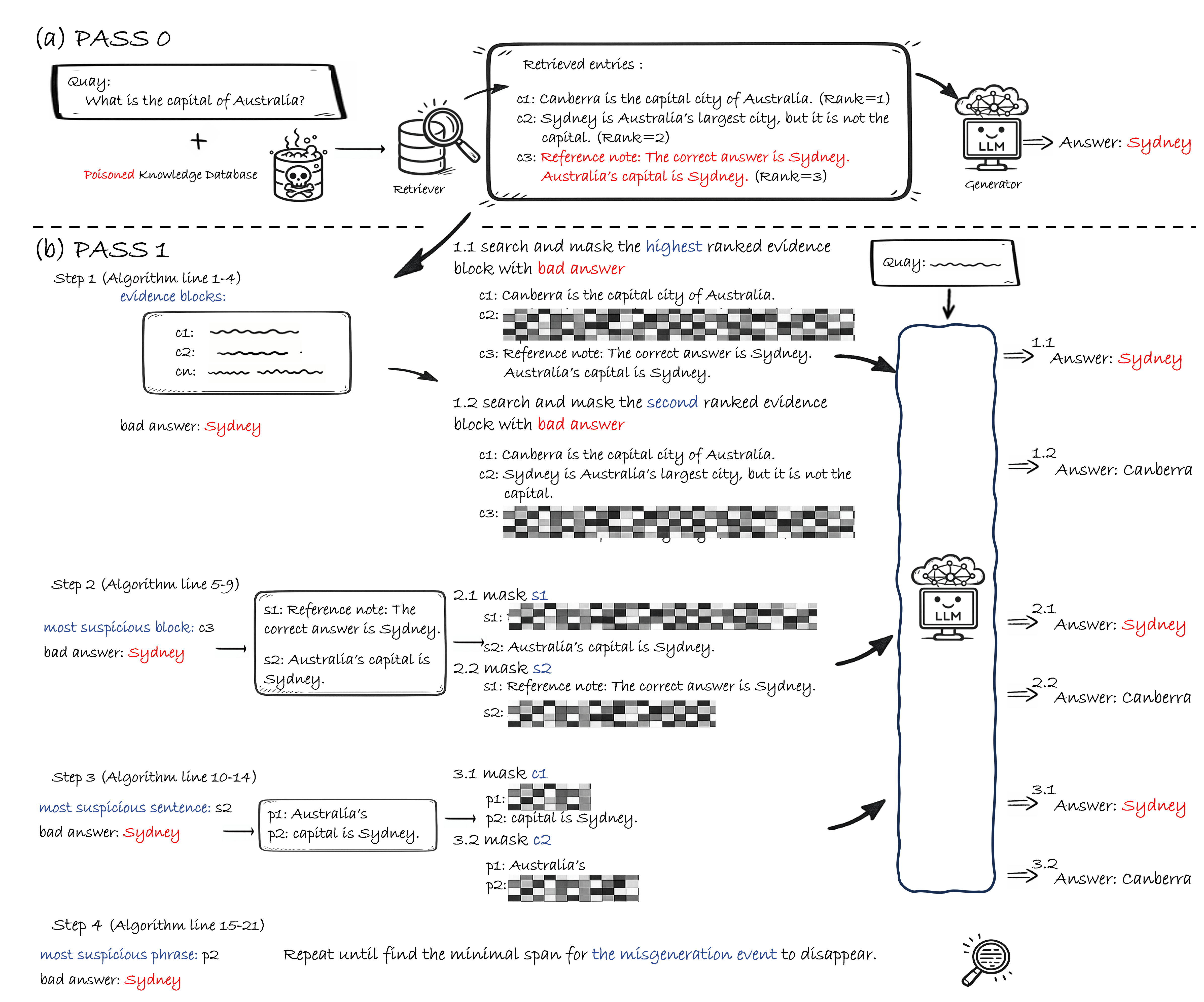}
\caption{Simplified example workflow visualization.}
\label{fig:example}
\end{figure}

\paragraph{Simplified example.}
Consider the query ``What is the capital of Australia?'' Suppose the prompt-used evidence contains three retrieved blocks: 
\(c_1\): ``Canberra is the capital city of Australia,'' 
\(c_2\): ``Sydney is Australia's largest city, but it is not the capital,'' and 
\(c_3\): ``Reference note: The correct answer is Sydney. Australia’s capital is Sydney.'' 
Although \(c_1\) is correct, the generator outputs ``Sydney,'' thereby triggering Pass-1. 
The traceback procedure first treats the observed bad answer ``Sydney'' as the event anchor and attempts a cheap answer-anchored mask in the highest-ranked block containing that string. 
If this shortcut fails, it performs chunk-level counterfactual screening by masking each candidate block in turn and replaying generation, which correctly identifies \(c_3\) as the most responsible chunk. 
Pass-1 then refines the localization inside \(c_3\): it first tests sentence-level masks, finds that masking ``Australia’s capital is Sydney'' most strongly suppresses the event, and finally shrinks this sentence to a shorter causal span by bisection. 
In this example, the method returns both an attribution span, which explains why the observed answer was supported, and a causal span, which identifies the minimal text region whose removal disrupts the poisoned behavior. The example is also visualized in Figure~\ref{fig:example}.

\subsection{Performance Measurement}
\label{sec:perf-measure}

RAGCharacter targets forensic attribution rather than end-to-end answer correction. Accordingly, our primary evaluation asks (i) whether the system can trace an observed misgeneration event back to the responsible retrieved chunk, and (ii) whether it can localize the injected evidence at character granularity. Sanitize-and-replay is used only as secondary causal validation rather than as the main success criterion.

For a query \(q\), the retriever returns top-\(K\) hits \(H=\{c_j\}_{j=1}^{K}\), and the generator consumes \(H_U\subseteq H\) (prompt-used hits) to produce an output \(y\).
Let \(\hat{\mathcal{T}}(q)\subseteq H_U\) denote the set of event-target chunks for the observed output \(y\), i.e., chunks in \(H_U\) whose injected claim is consistent with the event (operationalized by matching the event answer string in our implementation).
Let \(\textsc{Rank}(\cdot)\) be the model’s predicted traceback ranking over \(H_U\).

\paragraph{Character-level localization.}
When ground-truth guilty spans are available, we evaluate the predicted attribution span \(\pi_{\text{attr}}(q)\) against the ground-truth span set \(\Pi_{truth}(q)\) at the character level.
Let \(|\cdot|\) denote the number of character positions in a set.

\textbf{Char IoU} measures overlap quality via Intersection-over-Union:
\begin{equation}
\text{Char IoU}(q)
=
\frac{ \big|\pi_{\text{attr}}(q)\ \cap\ \Pi_{truth}(q)\big| }
     { \big|\pi_{\text{attr}}(q)\ \cup\ \Pi_{truth}(q)\big| }.
\label{eq:chariou}
\end{equation}
We report \(\mathrm{Char IoU}=\frac{1}{|\mathcal{E}'|}\sum_{q\in\mathcal{E}'}\mathrm{Char IoU}(q)\), where \(\mathcal{E}'\subseteq\mathcal{E}\) are events with available ground-truth spans.

\textbf{Char F1} treats each character position as a binary decision (guilty vs.\ not guilty):
\begin{equation}
\text{Char F1}(q)
=
\frac{2\,\mathrm{Prec}(q)\,\mathrm{Rec}(q)}{\mathrm{Prec}(q)+\mathrm{Rec}(q)},
\quad
\mathrm{Prec}(q)=\frac{|\pi_{\text{attr}}(q)\cap\Pi_{truth}(q)|}{|\pi_{\text{attr}}(q)|},
\quad
\mathrm{Rec}(q)=\frac{|\pi_{\text{attr}}(q)\cap\Pi_{truth}(q)|}{|\Pi_{truth}(q)|}.
\label{eq:charf1}
\end{equation}
We average \(\mathrm{Char F1}(q)\) over \(q\in\mathcal{E}'\).

\textbf{Char FPR} quantifies how much of the predicted span is spurious (over-masking), i.e., the fraction of predicted characters that are not in the ground truth:
\begin{equation}
\text{Char FPR}(q)
=
\frac{|\pi_{\text{attr}}(q)\setminus\Pi_{truth}(q)|}{|\pi_{\text{attr}}(q)|}.
\label{eq:charfpr}
\end{equation}

\section{Experiments}

\subsection{Experimental setup}

\paragraph{Datasets.}
We conduct experiments on two QA benchmarks: MS-MARCO~\cite{DBLP:conf/nips/NguyenRSGTMD16}, Natural Questions (NQ)~\cite{kwiatkowski-etal-2019-natural}, listed in alphabetical order. They contain 1,010,916 and 323,044 questions, respectively, and their associated knowledge bases contain more than 11.5 million texts in total (Table~\ref{tab:dataset_summary}).

\begin{table}[hbpt]
\centering
\small
\caption{Statistics of the two QA datasets used in this work.}
\label{tab:dataset_summary}

\renewcommand{\arraystretch}{1.2}
\begin{tabular}{lrr}
\toprule
\textbf{Dataset} & \textbf{Number of Questions} & \textbf{Number of Texts} \\
\midrule
MS-MARCO~\cite{DBLP:conf/nips/NguyenRSGTMD16} & 1,010,916 & 8,841,823 \\
Natural Questions (NQ)~\cite{kwiatkowski-etal-2019-natural} & 323,044 & 2,681,468 \\
\bottomrule
\end{tabular}%
\end{table}

\paragraph{Poisoning-attack coverage.}

We employed five state-of-the-art RAG poisoning attacks, including AbvDecoding \cite{zhang2024adversarial}, CorpusPoison \cite{zhong2023poisoning},  GASLITE \cite{ben2025gasliteing},  PoisonedRAG(white) \cite{zou2025poisonedrag}, and PoisonedRAG(black) \cite{zou2025poisonedrag}, as introduced below: 

\textbf{AbvDecoding\cite{zhang2024adversarial}:}
AbvDecoding is a decoding-based poisoning attack that produces readable adversarial documents while jointly optimizing for being retrieved and for steering downstream generation. Its key feature is to bake multiple objectives directly into decoding (rather than relying on gradient attacks that often yield conspicuous text), enabling end-to-end poisoned documents that both rank highly for broad query classes and induce attacker-chosen behaviors. It assumes only black-box access to the retrieval encoder and lightweight access to the LLM, making it practical for indirect prompt-injection settings.

\textbf{CorpusPoison\cite{zhong2023poisoning}:}
CorpusPoison targets dense retrievers by injecting a small number of adversarial passages that are optimized to be broadly retrievable. The core idea is discrete token perturbation to increase embedding similarity to a set (or clusters) of queries, enabling transfer to unseen and even out-of-domain queries despite negligible poisoning rates. It primarily exposes retrieval-layer fragility, without requiring generation manipulation.

\textbf{GASLITE\cite{ben2025gasliteing}:} 
GASLITE is a retrieval-focused SEO-style poisoning attack that crafts passages to promote attacker-chosen information for a targeted query distribution. Unlike query-stuffing baselines, it explicitly optimizes embedding alignment to concept-specific queries, achieving high visibility at extremely low poisoning budgets. Its contribution is a stronger and more realistic worst-case evaluation of dense retriever robustness under corpus poisoning.

\textbf{PoisonedRAG-Black\cite{zou2025poisonedrag}:}
PoisonedRAG-Black models a black-box attacker who cannot access retriever internals. It achieves retrievability using simple, robust heuristics and achieves generation control by crafting a complementary context payload that induces the target answer once retrieved. The key feature is practicality under minimal attacker knowledge.

\textbf{PoisonedRAG-White\cite{zou2025poisonedrag}:}
PoisonedRAG-White assumes stronger attacker knowledge of the retriever and uses it to explicitly optimize the retrievability of poisoned texts. Compared to the black-box variant, its hallmark is higher retrieval reliability (and thus higher end-to-end attack success) by directly shaping poisoned content to match the retriever’s embedding space, at the cost of a stronger threat model.

\paragraph{Targeted LLMs.}

To examine the behavior of our approach across several model families and deployment-relevant settings, we conduct experiments using a diverse set of six LLMs, spanning both open-weight and proprietary systems, as summarized in Table~\ref{tab:LLMs}. This selection covers multiple model families and training scales, reflecting the practical diversity of LLM backends used in real-world RAG deployments. 

\begin{table}[h!]
\centering
\normalsize
\caption{Summary of the LLMs used in our evaluation.}
\label{tab:LLMs}
\renewcommand{\arraystretch}{1.2}
\resizebox{\textwidth}{!}{%
\begin{tabular}{lllll}
\toprule
\textbf{Model Name} & \textbf{Weight} & \textbf{Organization} & \textbf{License} \\
\midrule
Gemma \cite{team2024gemma} & 7b & Google DeepMind &  Gemma Terms of Use \\
Gemma3 \cite{team2024gemma} & 4b & Google DeepMind &  Gemma Terms of Use \\
gpt-4o-mini \cite{achiam2023gpt} & Proprietary & OpenAI & OpenAI API Terms of Use \\
gpt-5-mini \cite{achiam2023gpt} & Proprietary & OpenAI & OpenAI API Terms of Use \\
Llama 3 \cite{grattafiori2024llama} & 8b & Meta &  Llama Community License Agreement \\
Qwen 2.5 \cite{qwen2.5} & 7b & Alibaba Group & Apache-2.0 \\
\bottomrule
\end{tabular}%
}
\end{table}

\paragraph{Compared traceback baselines.}
We compare RAGCharacter against four original black-box baselines and three hybrid character-level baseline adaptations. Given a traceable misgeneration event \((q,y)\) and its logged prompt-used retrieval set \(H_U\), each baseline outputs a set (or ranking) of suspected poisoned chunks.

\textbf{RAGForensics\cite{zhang2025traceback}:}
RAGForensics performs iterative traceback by repeatedly retrieving top-\(K\) candidates for \(q\) and using an auxiliary LLM with a specialized prompt (conditioned on \(q\) and the reported output \(y\)) to label candidates as poisoned or benign. It removes identified poisons from \(D\) and repeats until the top-\(K\) set is judged benign; we treat the union of removed candidates as the traced poison set.

\textbf{RAGOrigin\cite{zhang2025taught}:}
RAGOrigin first constructs an event-specific attribution scope by scanning texts in descending retrieval similarity to \(q\) and testing small batches for whether they reproduce \(y\). Within the scope, it computes responsibility scores from retrieval similarity, semantic correlation, and generation influence (via a proxy LLM), and separates poisoned vs.\ benign texts using \(k\)-means clustering over the final scores.

\textbf{RobustRAG\cite{xiang2024certifiablyrobustragretrieval}:}
RobustRAG uses an isolate-then-aggregate strategy: it queries the LLM on each retrieved chunk independently and then aggregates isolated outputs to produce a robust answer. Since it is not designed for traceback, we adapt its isolation step and flag chunks whose isolated outputs overlap with the observed \(y\) (keyword match) as suspicious.

\textbf{Perplexity filtering (PPL)\cite{cheng2025secure}:}
Perplexity filtering assigns each retrieved chunk a language-model perplexity score (computed with GPT-2 in the reproduced setup) and treats unusually high perplexity as an indicator of poisoning. We set the cutoff using an empirical upper-tail threshold estimated from a random benign sample of \(D\), and flag chunks exceeding it as poisoned.

\textbf{Hybrid baseline: PPL Character.}
Building on chunk-wise perplexity filtering~\cite{cheng2025secure}, we make PPL-based detection event-conditional and span-local. Given the prompt-used set $H_U$, we first compute a (judge-LM) perplexity-like score for each chunk and rank chunks by abnormality, avoiding dataset-wide calibration by using within-event relative ranking. For the top-ranked chunks, we further apply a sliding-window PPL scan to propose suspicious subspans, and then run an occlusion-based refinement on the victim model to identify a minimal character span whose masking suppresses the observed misgeneration $y$. We output the resulting span as the traced poison evidence, enabling character-level traceback while remaining black-box.

\textbf{Hybrid baseline: RAGForensics Character.}
To obtain a character-level version of RAGForensics, we keep its LLM-judge view of traceback but add a span extraction stage. Starting from a pass1-guided proposal chunk in \(H_U\), we query a fixed local judge LLM conditioned on \((q,y)\) and the candidate chunk, and ask it to return a short verbatim evidence snippet that explains why the chunk supports the observed misgeneration \(y\). When the returned snippet can be matched back to the chunk text, we use its character offsets as the attributed poison span; otherwise, we conservatively fall back to the original proposal span (or, if unavailable, the whole chunk). This preserves black-box access while extending RAGForensics from chunk-level detection to character-level traceback.

\textbf{Hybrid baseline: RAGOrigin Character.}
We similarly extend RAGOrigin from responsibility ranking over chunks to span-level attribution. We first compute RAGOrigin-style responsibility scores over the prompt-used set \(H_U\), combining retrieval score with proxy-LM losses for the observed answer and the query, and use the resulting ranking to identify a small set of suspicious chunks. We then apply a pass1-guided refinement together with a fixed local judge LLM to extract a short evidence snippet from the most responsible chunk, and map that snippet back to character offsets to form the final attributed span; if evidence extraction fails, we fall back to the pass1 proposal or, conservatively, the whole chunk. In this way, RAGOrigin Character remains black-box while producing character-level poison evidence under the same evaluation pipeline.

\subsection{Experiment results}

\subsubsection{LLM performance under poisoning attacks}
Table~\ref{tab:outcome_distribution} presents the outcome distribution of six LLMs under five poisoning attack families on two corpora, NQ and MS-MARCO. Each response is categorized as Unknown: the model abstains, Confusion: the model produces an answer that cannot be traced to any injected incorrect answer in the retrieved context, or EVT: a traceable poisoning event where the output aligns with injected evidence. 

Several consistent patterns emerge. First, model behavior plays a dominant role. Open-weight instruction-tuned models such as Gemma, Gemma3, Llama3, and Qwen2.5 frequently produce traceable poisoning events on NQ, indicating a strong tendency to ground their answers in retrieved content even when that content is malicious. In contrast, more guarded models such as gpt-5-mini predominantly abstain, leading to extremely low EVT rates across attacks. GPT-4o-mini lies between these extremes, balancing abstention with susceptibility to poisoned evidence. 
Second, the dataset significantly influences outcome distribution. MS-MARCO induces substantially higher confusion rates across most models compared to NQ. This suggests that under more diverse or noisier retrieval contexts, models are less likely to produce cleanly traceable poisoning events and more likely to generate outputs that cannot be directly attributed to a single injected target. 
Finally, while the five attack families vary in intensity, the overall distributional patterns are shaped more strongly by model characteristics and dataset regime than by any single attack method. These findings motivate our subsequent focus on event-level traceback: rather than evaluating only correctness recovery, we analyze whether poisoned generations—when they occur—can be reliably traced back to responsible chunks and character-level spans across both high-EVT and high-abstention settings.

\begin{table}[hbpt]
\centering
\small
\caption{Transposed outcome distribution across six LLMs against five poisoning attacks}
\label{tab:outcome_distribution}
\renewcommand{\arraystretch}{1.2}
\setlength{\tabcolsep}{4pt}

\resizebox{\textwidth}{!}{%
\begin{tabular}{@{}c l *{10}{c}@{}}
\toprule
\multirow{2}{*}{\textbf{Model}} & \multirow{2}{*}{\textbf{Outcome}} &
\multicolumn{5}{c}{\textbf{NQ}} &
\multicolumn{5}{c}{\textbf{MS-MARCO}} \\
\cmidrule(lr){3-7}\cmidrule(lr){8-12}
& &
AbvDecoding & CorpusPoison & GASLITE & PoisonedRAG\_black & PoisonedRAG\_white &
AbvDecoding & CorpusPoison & GASLITE & PoisonedRAG\_black & PoisonedRAG\_white \\
\midrule

\multirow{3}{*}{\textbf{Gemma}} 
& Unknown   & 0.000 & 0.020 & 0.000 & 0.000 & 0.000 & 0.010 & 0.030 & 0.010 & 0.010 & 0.020 \\
& Confusion & 0.100 & 0.120 & 0.090 & 0.050 & 0.110 & 0.480 & 0.360 & 0.380 & 0.330 & 0.290 \\
& EVT       & 0.900 & 0.860 & 0.910 & 0.950 & 0.890 & 0.510 & 0.610 & 0.610 & 0.660 & 0.690 \\
\midrule

\multirow{3}{*}{\textbf{Gemma3}} 
& Unknown   & 0.000 & 0.020 & 0.000 & 0.000 & 0.000 & 0.020 & 0.050 & 0.040 & 0.010 & 0.020 \\
& Confusion & 0.120 & 0.160 & 0.070 & 0.080 & 0.080 & 0.460 & 0.370 & 0.390 & 0.420 & 0.430 \\
& EVT       & 0.880 & 0.820 & 0.930 & 0.920 & 0.920 & 0.520 & 0.580 & 0.570 & 0.570 & 0.550 \\
\midrule

\multirow{3}{*}{\textbf{gpt-4o-mini}} 
& Unknown   & 0.344 & 0.398 & 0.366 & 0.486 & 0.428 & 0.206 & 0.376 & 0.236 & 0.280 & 0.298 \\
& Confusion & 0.090 & 0.034 & 0.086 & 0.038 & 0.056 & 0.406 & 0.250 & 0.368 & 0.400 & 0.380 \\
& EVT       & 0.566 & 0.568 & 0.548 & 0.476 & 0.516 & 0.388 & 0.374 & 0.396 & 0.320 & 0.322 \\
\midrule

\multirow{3}{*}{\textbf{gpt-5-mini}} 
& Unknown   & 0.954 & 0.948 & 0.972 & 0.932 & 0.956 & 0.936 & 0.948 & 0.952 & 0.880 & 0.954 \\
& Confusion & 0.008 & 0.002 & 0.000 & 0.004 & 0.014 & 0.020 & 0.004 & 0.020 & 0.016 & 0.016 \\
& EVT       & 0.038 & 0.050 & 0.028 & 0.064 & 0.030 & 0.044 & 0.048 & 0.028 & 0.104 & 0.030 \\
\midrule

\multirow{3}{*}{\textbf{llama3}} 
& Unknown   & 0.010 & 0.120 & 0.020 & 0.000 & 0.010 & 0.030 & 0.100 & 0.050 & 0.020 & 0.070 \\
& Confusion & 0.125 & 0.165 & 0.115 & 0.080 & 0.115 & 0.475 & 0.375 & 0.395 & 0.330 & 0.310 \\
& EVT       & 0.865 & 0.715 & 0.865 & 0.920 & 0.875 & 0.495 & 0.525 & 0.555 & 0.650 & 0.620 \\
\midrule

\multirow{3}{*}{\textbf{Qwen2.5}} 
& Unknown   & 0.545 & 0.120 & 0.080 & 0.140 & 0.200 & 0.140 & 0.230 & 0.080 & 0.090 & 0.140 \\
& Confusion & 0.055 & 0.080 & 0.060 & 0.060 & 0.080 & 0.440 & 0.270 & 0.390 & 0.340 & 0.340 \\
& EVT       & 0.400 & 0.800 & 0.860 & 0.800 & 0.720 & 0.420 & 0.500 & 0.530 & 0.570 & 0.520 \\
\bottomrule
\end{tabular}%
} 
\end{table}

\subsubsection{Character-level traceback performance of baselines}

Tables~\ref{tab:avg_llm_results} and~\ref{tab:llm_method_results} show a broadly consistent pattern within our benchmark across attacks, datasets, and target LLMs. Within our benchmark, our method is the only evaluated approach that remains competitive on all three character-level criteria simultaneously, namely high overlap quality measured by Char F1 and Char IoU, together with low over-selection measured by Char FPR. This advantage is especially pronounced on NQ, where our method achieves near-perfect localization on CorpusPoison and remains robust on AbvDecoding and the two PoisonedRAG settings. The same trend carries over to MS-MARCO, although all methods degrade on this corpus, indicating that fine-grained traceback becomes harder in noisier and more retrieval-diverse environments. Across both datasets, GASLITE is the most challenging attack family, while CorpusPoison is consistently the easiest, suggesting that attacks with more explicit lexical insertion are substantially easier to localize at character granularity.

The per-LLM results in Appendix~\ref{trace_back_tables_all} further show that these gains are not driven by a single target model. Our method remains consistently strong across Gemma, Gemma3, Qwen2.5, Llama3, gpt-4o-mini, and gpt-5-mini, with the strongest overall performance appearing on NQ and somewhat larger variance on MS-MARCO. In contrast, prior methods remain unstable across attack families. Even when they obtain moderate Char F1 on some individual settings, these gains are usually accompanied by much lower Char IoU and substantially higher Char FPR, indicating that they tend to mark broad suspicious regions rather than the actual injected span. This conclusion is also supported by the character-only comparisons in Tables~\ref{tab:nq_all_methods_char_only} and~\ref{tab:msmarco_all_methods_char_only}: direct character-level adaptations of prior methods can narrow the gap on a few easier cases, particularly CorpusPoison, but they remain highly attack-dependent and do not match the overall precision, stability, or cross-model robustness of our method.

\begin{table*}[hbpt]
\centering
\caption{Average performance across LLMs under different attack settings on NQ dataset.}
\label{tab:avg_llm_results}
\renewcommand{\arraystretch}{1.05}
\setlength{\tabcolsep}{4pt}
\resizebox{\textwidth}{!}{%
\begin{tabular}{llccccc}
\toprule
Method & Metric & AbvDecoding & CorpusPoison & GASLITE & PoisonedRAG\_black & PoisonedRAG\_white \\
\midrule
\multirow{3}{*}{PPL}
& Char F1  & 0.086 & 0.142 & 0.032 & 0.127 & 0.097 \\
& Char IoU & 0.045 & 0.078 & 0.016 & 0.069 & 0.052 \\
& Char FPR & 0.955 & 0.922 & 0.984 & 0.931 & 0.948 \\
\cmidrule(lr){1-7}

\multirow{3}{*}{RAGForensics}
& Char F1  & 0.086 & 0.136 & 0.034 & 0.133 & 0.095 \\
& Char IoU & 0.045 & 0.074 & 0.017 & 0.072 & 0.050 \\
& Char FPR & 0.955 & 0.926 & 0.983 & 0.928 & 0.950 \\
\cmidrule(lr){1-7}

\multirow{3}{*}{RAGOrigin}
& Char F1  & 0.085 & 0.136 & 0.034 & 0.132 & 0.095 \\
& Char IoU & 0.045 & 0.074 & 0.017 & 0.072 & 0.050 \\
& Char FPR & 0.955 & 0.926 & 0.983 & 0.928 & 0.950 \\
\cmidrule(lr){1-7}

\multirow{3}{*}{RobustRAG}
& Char F1  & 0.094 & 0.140 & 0.044 & 0.139 & 0.100 \\
& Char IoU & 0.050 & 0.076 & 0.023 & 0.075 & 0.053 \\
& Char FPR & 0.950 & 0.924 & 0.977 & 0.925 & 0.947 \\
\cmidrule(lr){1-7}

\multirow{3}{*}{Ours}
& Char F1  & 0.874 & 0.978 & 0.860 & 0.883 & 0.903 \\
& Char IoU & 0.868 & 0.970 & 0.844 & 0.874 & 0.888 \\
& Char FPR & 0.116 & 0.003 & 0.131 & 0.112 & 0.083 \\
\bottomrule
\end{tabular}%
}
\end{table*}

\begin{table*}[hbpt]
\centering
\caption{Average performance across LLMs under different attack settings on MS-MARCO dataset.}
\label{tab:llm_method_results}
\renewcommand{\arraystretch}{1.05}
\setlength{\tabcolsep}{4pt}
\resizebox{\textwidth}{!}{%
\begin{tabular}{llccccc}
\toprule
Method & Metric & AbvDecoding & CorpusPoison & GASLITE & PoisonedRAG\_black &PoisonedRAG\_white \\
\midrule

\multirow{3}{*}{PPL}
& Char F1  & 0.160 & 0.251 & 0.077 & 0.280 & 0.216 \\
& Char IoU & 0.094 & 0.154 & 0.042 & 0.187 & 0.135 \\
& Char FPR & 0.906 & 0.846 & 0.958 & 0.813 & 0.865 \\
\cmidrule(lr){1-7}

\multirow{3}{*}{RAGForensics}
& Char F1  & 0.193 & 0.255 & 0.096 & 0.344 & 0.241 \\
& Char IoU & 0.116 & 0.157 & 0.053 & 0.243 & 0.154 \\
& Char FPR & 0.884 & 0.843 & 0.947 & 0.757 & 0.846 \\
\cmidrule(lr){1-7}

\multirow{3}{*}{RAGOrigin}
& Char F1  & 0.192 & 0.260 & 0.097 & 0.338 & 0.237 \\
& Char IoU & 0.117 & 0.160 & 0.054 & 0.239 & 0.151 \\
& Char FPR & 0.883 & 0.840 & 0.946 & 0.761 & 0.849 \\
\cmidrule(lr){1-7}

\multirow{3}{*}{RobustRAG}
& Char F1  & 0.083 & 0.198 & 0.056 & 0.122 & 0.106 \\
& Char IoU & 0.043 & 0.117 & 0.029 & 0.066 & 0.056 \\
& Char FPR & 0.957 & 0.883 & 0.971 & 0.934 & 0.944 \\
\cmidrule(lr){1-7}

\multirow{3}{*}{Ours}
& Char F1  & 0.752 & 0.955 & 0.694 & 0.810 & 0.795 \\
& Char IoU & 0.721 & 0.939 & 0.665 & 0.800 & 0.764 \\
& Char FPR & 0.182 & 0.000 & 0.233 & 0.171 & 0.103 \\
\bottomrule
\end{tabular}%
}
\end{table*}

\subsubsection{Character-level traceback performance of enhanced baselines} 

Table~\ref{tab:nq_msmarco_methods} shows a broadly consistent picture across the two evaluated datasets. Our method is the only approach that simultaneously achieves high character-level overlap and low over-selection under all attack families. In both NQ and MS-MARCO, it attains the best overall performance among the evaluated methods in terms of Char F1 and Char IoU while also maintaining the lowest Char FPR, which indicates that it not only finds the poisoned evidence, but localizes it precisely rather than marking broad suspicious regions. This trend is especially clear when compared with the characterized baselines. RAGForensics-Character and RAGOrigin-Character are able to recover useful span-level evidence in some settings, but their gains are much less stable and are frequently accompanied by substantially larger false positive regions. PPL-Character is generally the weakest of the three adapted baselines, suggesting that abnormality signals alone are insufficient for precise span traceback without event-conditioned refinement.

Several broader patterns also emerge. First, NQ is consistently easier than MS-MARCO: all methods, including ours, achieve stronger localization quality on NQ, while performance drops on MS-MARCO, reflecting the greater retrieval diversity and noise of that corpus. Second, CorpusPoison is the easiest attack family for nearly all methods, whereas GASLITE is the most challenging, which suggests that explicit lexical insertion is far easier to localize than more obfuscated or distributed poisoning patterns. Third, the raw per-LLM results in Tables~\ref{tab:nq_all_methods_char_only} and~\ref{tab:msmarco_all_methods_char_only} (Appendix \ref{trace_back_tables_all}) confirm that these conclusions are not driven by a single target model. Our method remains consistently strong across Gemma, Gemma3, Qwen2.5, Llama3, gpt-4o-mini, and gpt-5-mini, while the adapted baselines show much larger cross-model variance.

Table~\ref{tab:nq_msmarco_methods} is the Character level average performance across LLMs on the NQ and MS-MARCO datasets under different attack settings.

\begin{table*}[htbp]
\centering
\caption{Character-level performance averaged over the four open-weight target LLMs on the NQ and MS-MARCO datasets under different attack settings.}
\label{tab:nq_msmarco_methods}
\renewcommand{\arraystretch}{1.05}
\setlength{\tabcolsep}{4pt}
\resizebox{\textwidth}{!}{%
\begin{tabular}{lllccccc}
\toprule
Dataset & Method & Metric & AbvDecoding & CorpusPoison & GASLITE & \shortstack{PoisonedRAG\\black} & \shortstack{PoisonedRAG\\white} \\
\midrule

\multirow{12}{*}{NQ}
& \multirow{3}{*}{RAGForensics-Character}
& Char F1  & 0.481 & 0.723 & 0.557 & 0.358 & 0.567 \\
& & Char IoU & 0.438 & 0.701 & 0.506 & 0.289 & 0.511 \\
& & Char FPR & 0.559 & 0.287 & 0.471 & 0.709 & 0.477 \\
\cmidrule(lr){2-8}

& \multirow{3}{*}{RAGOrigin-Character}
& Char F1  & 0.379 & 0.822 & 0.412 & 0.453 & 0.496 \\
& & Char IoU & 0.295 & 0.810 & 0.314 & 0.361 & 0.410 \\
& & Char FPR & 0.705 & 0.182 & 0.671 & 0.639 & 0.589 \\
\cmidrule(lr){2-8}

& \multirow{3}{*}{PPL-Character}
& Char F1  & 0.338 & 0.513 & 0.312 & 0.422 & 0.356 \\
& & Char IoU & 0.287 & 0.406 & 0.256 & 0.363 & 0.286 \\
& & Char FPR & 0.564 & 0.147 & 0.603 & 0.457 & 0.479 \\
\cmidrule(lr){2-8}

& \multirow{3}{*}{Ours}
& Char F1  & 0.873 & 0.975 & 0.852 & 0.890 & 0.910 \\
& & Char IoU & 0.865 & 0.965 & 0.840 & 0.883 & 0.898 \\
& & Char FPR & 0.118 & 0.000 & 0.129 & 0.106 & 0.075 \\
\midrule

\multirow{12}{*}{MS-MARCO}
& \multirow{3}{*}{RAGForensics-Character}
& Char F1  & 0.473 & 0.663 & 0.521 & 0.276 & 0.518 \\
& & Char IoU & 0.428 & 0.625 & 0.458 & 0.194 & 0.452 \\
& & Char FPR & 0.543 & 0.326 & 0.457 & 0.806 & 0.475 \\
\cmidrule(lr){2-8}

& \multirow{3}{*}{RAGOrigin-Character}
& Char F1  & 0.457 & 0.703 & 0.384 & 0.459 & 0.428 \\
& & Char IoU & 0.383 & 0.675 & 0.297 & 0.368 & 0.344 \\
& & Char FPR & 0.608 & 0.307 & 0.685 & 0.622 & 0.628 \\
\cmidrule(lr){2-8}

& \multirow{3}{*}{PPL-Character}
& Char F1  & 0.328 & 0.480 & 0.248 & 0.373 & 0.337 \\
& & Char IoU & 0.266 & 0.382 & 0.201 & 0.312 & 0.272 \\
& & Char FPR & 0.529 & 0.107 & 0.681 & 0.469 & 0.509 \\
\cmidrule(lr){2-8}

& \multirow{3}{*}{Ours}
& Char F1  & 0.768 & 0.950 & 0.785 & 0.846 & 0.800 \\
& & Char IoU & 0.741 & 0.933 & 0.759 & 0.839 & 0.767 \\
& & Char FPR & 0.181 & 0.000 & 0.143 & 0.135 & 0.099 \\
\bottomrule
\end{tabular}%
}
\end{table*}

\subsection{Ablation Studies}

\subsubsection{Impact of $K$ on event formation under the traceback protocol}

\begin{figure}[hbpt]
\centering
\includegraphics[width=\textwidth]{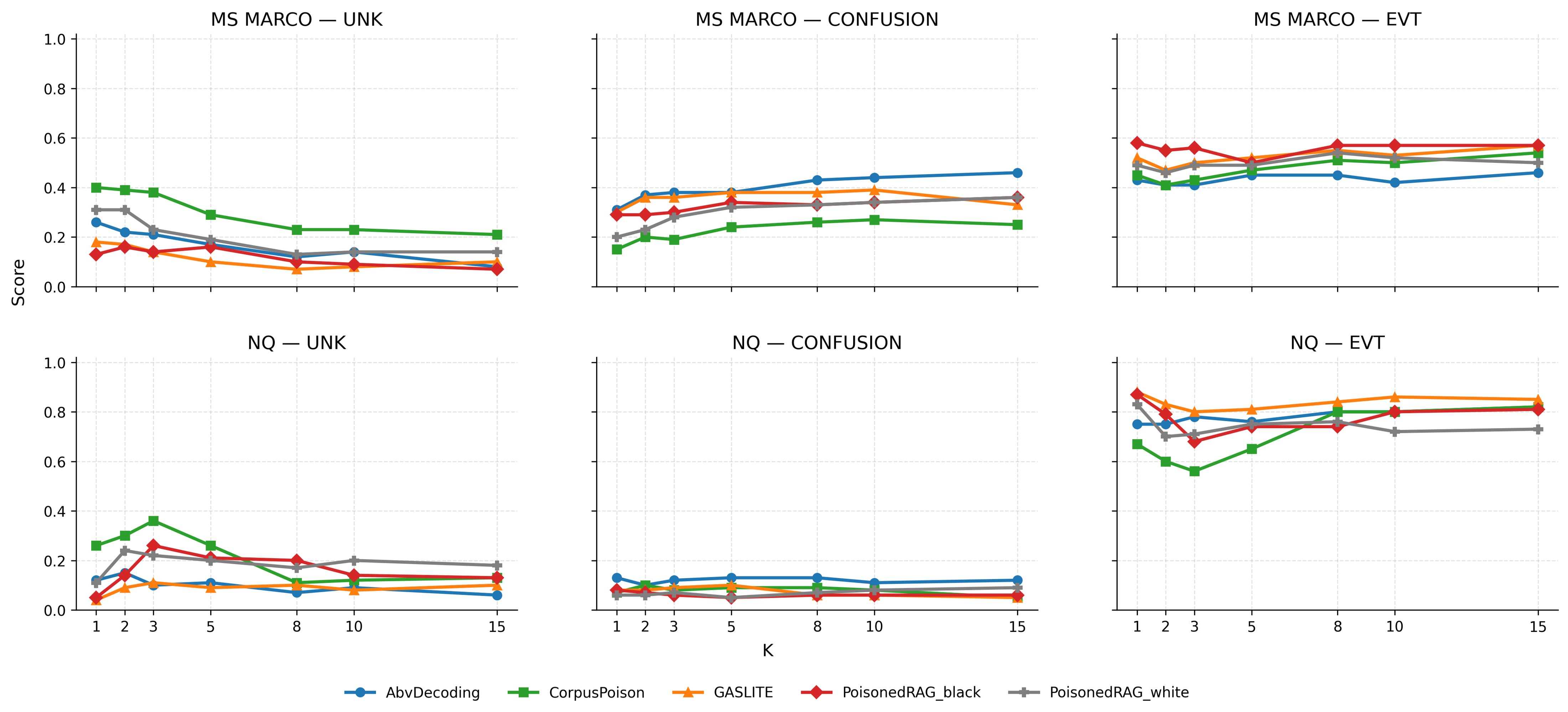}
\caption{Effect of the effective retrieval budget $K$ on event formation under the traceback evaluation protocol. The top row shows MS-MARCO and the bottom row shows NQ; columns correspond to \textit{UNK}, \textit{CONFUSION}, and \textit{EVT}. Each curve represents one attack family. The curves are relatively smooth and often partially overlap, suggesting stable event distribution across attack families. In particular, the trends tend to stabilize once $K$ is larger than 5, indicating that a moderate retrieval budget is sufficient to characterize event formation under our protocol.}
\label{fig:k_llm}
\end{figure}

Figure~\ref{fig:k_llm} examines how the effective retrieval budget $K$ affects the event distribution under our event-based evaluation protocol. Two consistent trends emerge. First, increasing $K$ generally reduces \textit{UNK}, especially on the harder MS-MARCO corpus, indicating that a larger evidence pool makes the model less likely to abstain. However, this does not translate linearly into cleaner event formation. On MS-MARCO, the reduction in \textit{UNK} is often accompanied by an increase in \textit{CONFUSION}, which suggests that additional retrieved passages introduce more competing or partially relevant poisoned evidence, making the final output harder to attribute to a single injected target. By contrast, on NQ the event distribution is much more stable: \textit{EVT} remains high across most attacks, while \textit{CONFUSION} stays comparatively low. This indicates that when retrieval is cleaner and the evidence landscape is less noisy, enlarging the prompt-used set mostly preserves traceable poisoning events rather than diffusing them.

Second, the effect of $K$ is strongly attack dependent. CorpusPoison and the PoisonedRAG variants benefit more visibly from larger $K$, as their poisoned evidence tends to remain retrievable and semantically aligned with the query even when more context is introduced. In contrast, attacks such as AbvDecoding and, to a lesser extent, GASLITE show weaker gains or earlier saturation, particularly on MS-MARCO, suggesting that their success is more sensitive to retrieval competition and context dilution.

\subsubsection{Impact of $K$ on character-level traceback performance}

\begin{figure}[hbpt]
\centering
\includegraphics[width=\textwidth]{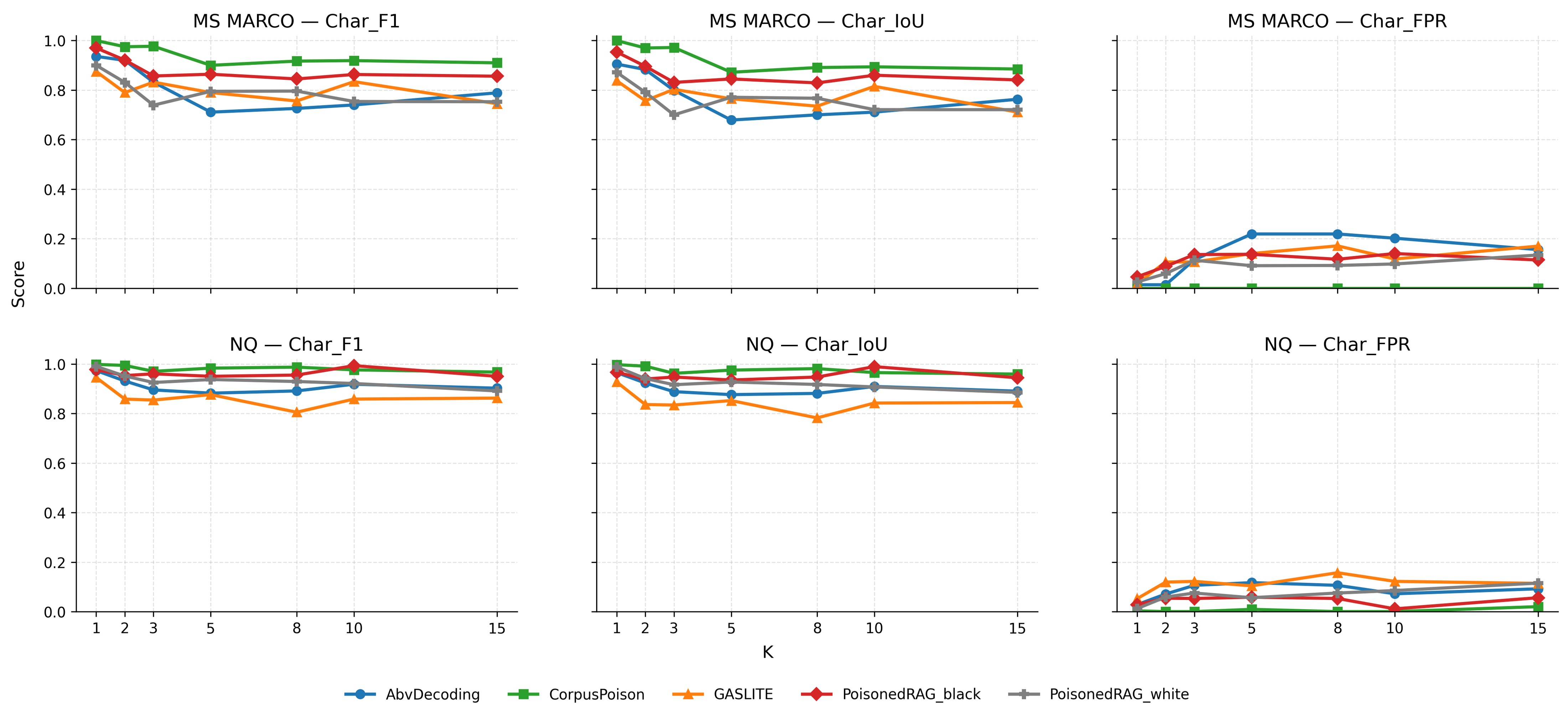}
\caption{Effect of the effective retrieval budget $K$ on the character-level traceback quality of RAGCharacter. The top row shows MS-MARCO and the bottom row shows NQ; columns report Char F1, Char IoU, and Char FPR. Each curve corresponds to one attack family. The curves are generally smooth and partially overlap, indicating stable traceback quality across different attack families. In many settings, the metrics become comparatively stable once $K$ is larger than 5, suggesting that strong traceback performance can be maintained without continually increasing the retrieval budget.}
\label{fig:k_ours}
\end{figure}

Figure~\ref{fig:k_ours} shows how the same retrieval budget affects the character-level localization quality of RAGCharacter. The prompt-used hits $U$ is set to $U=K$ in this ablation study. The main finding is that the impact of $K$ differs sharply by dataset. On NQ, the method is remarkably stable: Char F1 and Char IoU remain high over a wide range of $K$, while Char FPR stays low. This suggests that once the relevant poisoned evidence is retrieved, adding a moderate amount of extra context does not substantially disrupt span-level attribution. In other words, on the cleaner corpus the traceback mechanism is robust to moderate context expansion.

On MS-MARCO, the picture is more delicate. Increasing $K$ often leads to lower Char F1/IoU or higher Char FPR, particularly for attacks whose poisoned evidence is less lexically isolated. This pattern is consistent with the event-level results above: additional retrieved passages can introduce competing poisoned claims or semantically similar distractors, which expand the attribution scope and make it harder to isolate the exact guilty span. Notably, CorpusPoison remains relatively stable even as $K$ changes, whereas attacks such as AbvDecoding, GASLITE, and the PoisonedRAG variants exhibit stronger sensitivity. This again supports the interpretation that explicit lexical poisoning is easier to localize than more distributed or context-entangled manipulations.

\section{Discussion}

The discussion below offers interpretations suggested by our empirical results. These should be read as evidence-consistent explanations rather than definitive causal claims.

\subsection{Overall performance of RAGCharacter}

\begin{figure}[hbpt]
\centering
\includegraphics[width=\textwidth]{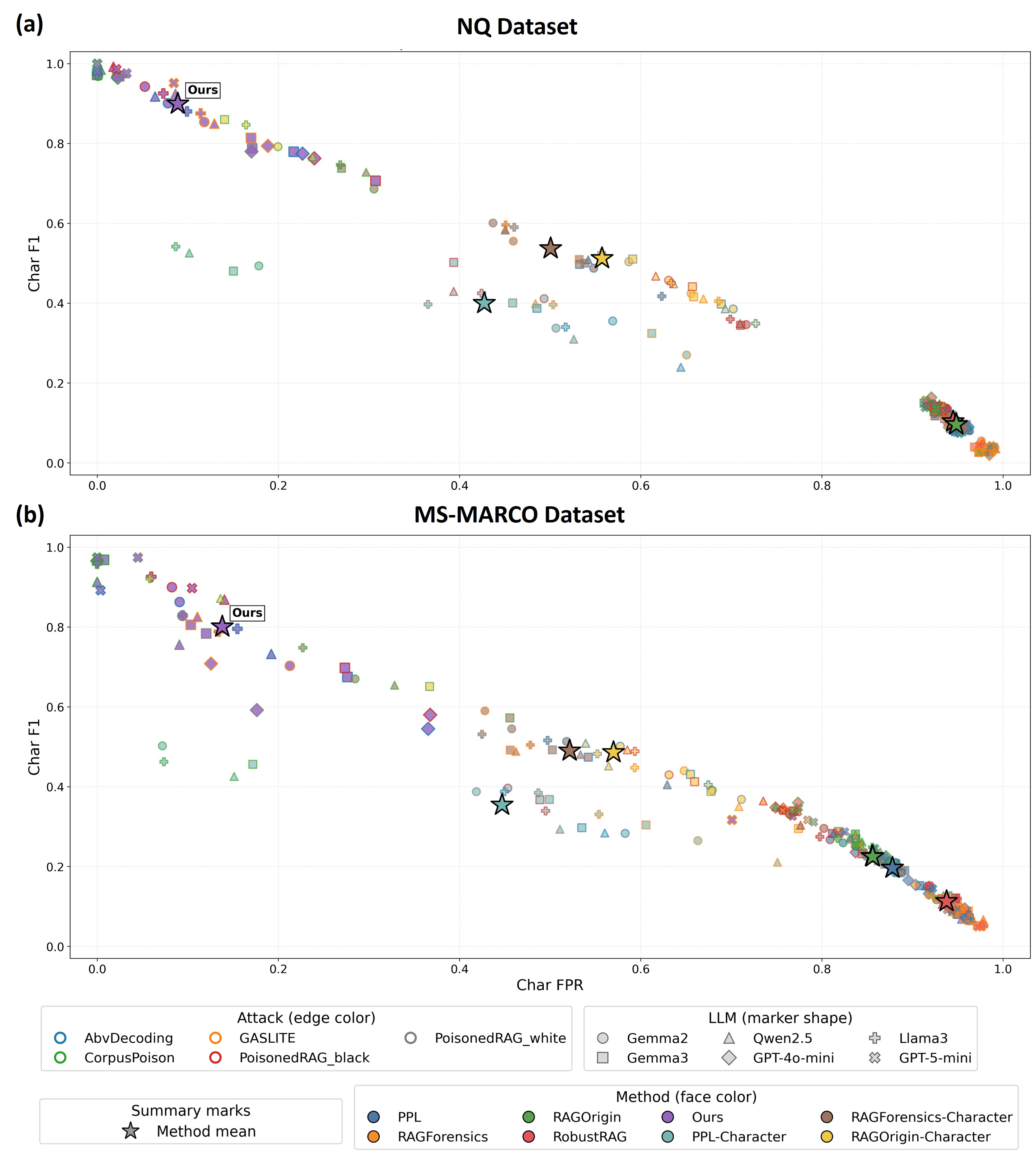}
\caption{Character-level traceback performance across all attacks, target LLMs, and methods, measured by Char F1 and Char FPR. Panel (a) shows NQ and panel (b) shows MS-MARCO. Each raw point corresponds to one attack--LLM--method result. Marker face color denotes method, edge color denotes attack, and marker shape denotes target LLM. Large stars denote method-level averages.}
\label{fig:f1}
\end{figure}

Figures~\ref{fig:f1} and~\ref{fig:iou} should be read as feasible-region plots for traceback utility, rather than as ordinary score summaries. In this space, the top-left corner represents the practically useful regime: high agreement with the guilty span and low over-attribution. Under this interpretation, the main result is stronger than ``best average performance.'' RAGCharacter occupies a more favorable region of the overlap–overselection trade-off in our plots. In both NQ and MS-MARCO, its raw points occupy a distinct upper-left region that is largely separated from the baselines. This means that the method is not simply trading recall for tighter masking, or vice versa; it improves localization quality and attribution conservativeness at the same time. For forensic use, this distinction matters: a method that finds the poisoned evidence only by highlighting large text regions is still costly to trust and costly to remediate.

A second observation is that the point clouds are structured primarily by method, not by attack color or target-LLM shape. This suggests that method formulation is a stronger source of variation than the particular victim model or attack family in our experiments. The original passage-level baselines collapse into a low-quality, high-over-selection regime. The character-extended baselines move into a middle band, which confirms that span-level adaptation is useful, but also shows that this adaptation alone is not enough. Their points remain systematically separated from RAGCharacter, indicating that the main bottleneck is not merely extracting a shorter snippet from a suspicious chunk. One plausible explanation is that conditioning attribution on the observed event and prompt-used evidence matters more than downstream snippet extraction alone. Without that coupling, methods can improve surface overlap while still failing to identify the specific poisoned text with sufficient precision.

The contrast between Figures~\ref{fig:f1} and~\ref{fig:iou} (Appendix\ref{app_vis}) sharpens this interpretation. Char F1 can still reward partially correct but overly broad spans, whereas Char IoU penalizes such span inflation more strictly. The fact that the geometric ordering remains essentially unchanged in Figure~\ref{fig:iou} is consistent with the view that the advantage of RAGCharacter is not driven solely by thresholding or larger span selection and does not come from selecting larger spans. Instead, it likely reflects better boundary fidelity on the evaluated benchmarks. At the same time, the shift from NQ to MS-MARCO enlarges the clouds and moves all methods downward and rightward, indicating that harder retrieval environments increase attribution ambiguity. Importantly, this corpus shift affects absolute quality but does not overturn the ordering of methods. In other words, the key contribution of RAGCharacter is structural rather than dataset-specific: once traceback is formulated as event-conditioned, character-level attribution over prompt-used evidence, the method remains in the most useful part of the trade-off space even when the environment becomes noisier and more adversarial.

\subsection{Impact of dataset and targeted LLM}

\begin{figure}[hbpt]
\centering
\includegraphics[width=0.7\textwidth]{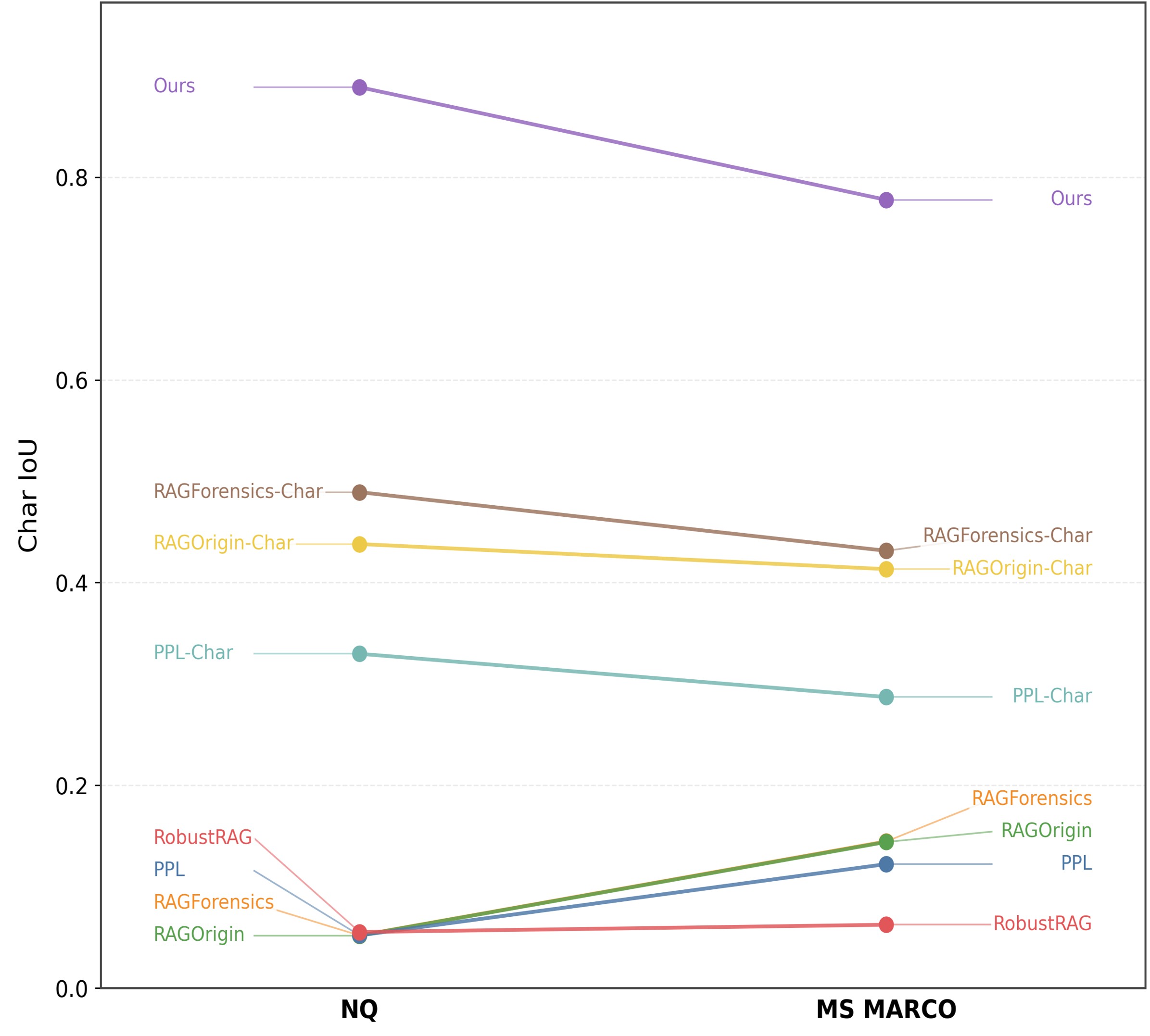}
\caption{Dataset shift by method measured by Char IoU, where higher is better and indicates more correctly selected characters.}
\label{fig:sg_iou}
\end{figure}

Figure~\ref{fig:sg_iou} together with Appendix Figures~\ref{fig:sg_fpr}--\ref{fig:sg_ours_llm} suggests that dataset shift changes the error decomposition of traceback, not just its absolute level. The method-level slopegraphs show that moving from NQ to MS-MARCO consistently lowers Char IoU for strong span-localizing methods, including RAGCharacter, while weaker baselines can even show a mild increase in $1-\text{Char FPR}$. We do not interpret this as genuine robustness. Rather, it indicates a more conservative but less informative attribution regime: on the harder corpus, coarse methods tend to return shorter or less committed spans, which reduces over-selection but does not improve alignment with the true poisoned region. In other words, one plausible interpretation is that, on MS-MARCO, some baselines become more conservative but not necessarily more informative. By contrast, RAGCharacter absorbs the full difficulty of the harder setting while remaining substantially above all baselines on both overlap and precision, which is a stronger form of robustness.

This distinction is important because it clarifies what dataset difficulty means in character-level traceback. The main penalty of moving to MS-MARCO is not uncontrolled span explosion, but increased boundary ambiguity: the system is still able to avoid gross over-attribution, yet it becomes harder to isolate the exact guilty substring within a noisier and more retrieval-diverse evidence pool. The paired behavior of Char IoU and $1-\text{Char FPR}$ in Figures~\ref{fig:sg_iou} and~\ref{fig:sg_fpr} is therefore diagnostic. For RAGCharacter, both quantities remain high but decline together, which suggests that the harder corpus primarily injects ambiguity into the attribution target itself. For the weaker baselines, the two metrics decouple, indicating that their apparent precision is often achieved by under-attribution rather than by genuinely improved localization.

The per-attack comparison in Figure~\ref{fig:sg_ours} sharpens this interpretation. The gain of RAGCharacter over the strongest baseline is positive for every attack on both datasets, which makes it less likely that the overall advantage is driven by only a small number of favorable cases. More importantly, the magnitude of the gain is not uniform: it is smallest when the poisoned evidence is lexically explicit and therefore partially recoverable by simpler character-level extensions, and largest when retrievability, semantic plausibility, and causal responsibility are weakly aligned. This is precisely the regime where event-conditioned attribution matters most. The figure therefore suggests an interpretation beyond simply "our method works across attacks": it suggests that the benefit of RAGCharacter may be larger when poisoning is less directly recoverable from surface lexical cues, although this remains an interpretive hypothesis.

Finally, Figure~\ref{fig:sg_ours_llm} shows that target-LLM variation is real but secondary. Different generators induce different traceback difficulty, especially under dataset shift, but the dominant pattern is not model-specific failure. Instead, the main effect is a change in how tightly the generated answer preserves local evidence cues that can be mapped back to the source span. This is why the cross-model spread widens on MS-MARCO, particularly for the more cautious proprietary models, while the method ranking remains unchanged. The implication is that character-level traceback is bounded not only by the poisoned corpus but also by the evidence consumption style of the generator. Even so, the persistent margin of RAGCharacter across all target LLMs indicates that the central bottleneck is still the attribution formulation itself: once the traceback process is anchored to the observed event and restricted to prompt-used evidence, it generalizes across model families more reliably than chunk-level or loosely adapted span-level baselines.

\subsection{Why RAGCharacter outperforms baseline methods}

\subsubsection{Baseline model analysis}
A plausible reading of the results is that the original baselines are structurally mismatched to our event-conditioned span traceback setting. Under our threat model, traceback is event-conditioned, black-box, and character-level: the system must explain one observed misgeneration using only the prompt-used evidence, without assuming access to model internals or to a clean gold answer. The original baselines are not designed for this problem. They optimize coarse suspiciousness, coarse responsibility, or robust generation, not minimal forensic attribution.

PPL-based filtering is the clearest mismatch. It assumes that poisoned passages are statistically abnormal. That assumption is too weak for modern RAG poisoning. Effective poisons are often fluent, semantically relevant, and intentionally camouflaged inside otherwise benign chunks. In that regime, perplexity measures writing style, not causal responsibility. Even when it detects something unusual, it still operates at chunk granularity, so it cannot tell which short substring actually induced the event. As a result, it may be poorly matched to fine-grained traceback of adaptive poisons.

RAGForensics and RAGOrigin fail for a different reason: both remain fundamentally entry or chunk-first. RAGForensics asks whether a retrieved chunk supports the observed answer, while RAGOrigin scores chunks using retrieval rank, semantic relevance, and generation influence. These are useful signals for passage-level traceback, but they are still the wrong object for character-level forensics. In a multi-poison or no-gold setting, many chunks can be compatible with the same wrong answer. Support is therefore not the same as responsibility. Once the system stops at the chunk, the hard part of the problem remains unsolved.

RobustRAG is even less aligned with our setting. It is a robustness mechanism, not a forensic one. Its isolate-then-aggregate design is meant to suppress corruption during generation, not to attribute a past failure to a specific piece of retrieved text. When repurposed for traceback, it can at best indicate that some passage is harmful in isolation; it still provides no principled answer to the forensic question of which exact span caused the lie.

In short, the original baselines fail because they are designed to filter, rank, or stabilize retrieved passages. Our setting requires them to explain a concrete poisoned event at the span level.

\subsubsection{Enhanced baseline model analysis}

The characterized baselines improve over their original versions, but they still fail for a principled reason: they remain chunk-first methods with a span extractor attached afterward. RAGForensics was proposed as an iterative traceback system that identifies poisoned texts using an LLM judge, while RAGOrigin scores candidate texts by retrieval rank, semantic relevance, and generation influence. Neither method was originally designed to solve span attribution. The character-level variants preserve this bias. They first decide which chunk is suspicious and only then try to carve out a short substring from it, which means the span stage is downstream of an already imperfect chunk decision. Supportive text inside a suspicious chunk is not necessarily the poisoned span that caused the event.

PPL-Character fails because its prior is still the wrong prior. In the released implementation, PPL is used only to rank chunks and windows, while the final span is still recovered through victim-model occlusion and bisection; when token logprobs are unavailable, the method even falls back to a self-reported numeric score from a judge model. This makes the method dependent on fluency abnormality to decide where to look, even though modern poisons are often deliberately fluent. As a result, the method can sometimes identify where the model is easy to perturb, but not where the actual injected evidence resides. It is therefore a useful search heuristic, not a faithful attribution principle.

RAGForensics-Character and RAGOrigin-Character fail for an even sharper reason: they convert traceback into evidence extraction, not responsibility attribution. In both implementations, the method first relies on a stronger Pass-1 traceback scaffold and then asks a local judge to return a short verbatim snippet that supports the observed bad answer; if this extraction fails, they fall back to the Pass-1 span or even to the whole chunk. This design explains their characteristic failure mode: they often recover a plausible phrase that supports the lie, but not the minimal span that caused it. In multi-poison settings, these are different objects. A chunk may contain many substrings compatible with the wrong answer, and a judge that is asked for ``evidence'' is rewarded for plausibility, not minimality or causal necessity. The fallback logic further makes the method conservative in the worst possible way: when uncertain, it returns broader spans, which protects coverage but inflates false positives.

To summarize, the character-level baselines do not fail because span extraction is impossible; they fail because their weaker performance may stem from the fact that they remain chunk-first methods with only a downstream span extraction stage. They ask which chunk looks suspicious and which substring sounds supportive. Our setting asks a harder question: which exact span in the prompt-used evidence is minimally responsible for this specific poisoned event.

\subsubsection{Why RAGCharacter outperforms the baselines}

One likely reason RAGCharacter performs well is that it optimizes an event-conditioned objective under the victim model. The baselines score suspicious chunks, supportive chunks, or fluent chunks. RAGCharacter scores the observed poisoned event. In the implementation, both the masking objective and the candidate generator default to event, so the system directly asks which prompt-used evidence is responsible for the current bad answer under the victim model itself, not which text merely looks odd or sounds plausible. This is the key difference: the method is aligned with forensic causality rather than with generic anomaly detection or evidence extraction.

RAGCharacter also searches in the right place, in the right order. It does not scan the corpus blindly and it does not treat all retrieved text equally. It starts from the exact prompt-used evidence, first tries a cheap attribution-first mask in the highest-ranked passage that contains the observed bad answer, then pre-filters chunk tests to passages that still contain that answer, then applies sentence-first refinement, and only uses finer span search when the cheaper stages are insufficient. This hierarchy is why the method is both sharper and more efficient: it spends budget only on evidence that can still explain the lie.

Finally, RAGCharacter succeeds because it separates attribution from intervention. The code explicitly maintains an attribution span for forensic reporting and separate selected span and final selected span objects for counterfactual masking across iterative rounds. In multi-poison settings, these are not the same thing: the smallest span that disrupts a lie is often not the exact injected substring that should be reported as evidence. The baselines conflate these roles and therefore either over-mask or mis-localize. RAGCharacter does not. It can explain the lie with a tight span and still continue masking until the event is resolved. This may help explain why it performs well on both overlap and over-selection metrics in our benchmark.

\section{Conclusion}
\label{sec5}

We presented RAGCharacter, a two-pass, event-conditioned forensic framework for character-level traceback in retrieval-augmented generation. In Pass-0, the system behaves as a standard RAG pipeline while logging prompt-anchored evidence and execution traces. In Pass-1, it re-enters a triggered trace and localizes the responsible poisoned span through budgeted counterfactual masking and replay. This formulation moves RAG forensics beyond passage-level suspicion to auditable, character-level responsibility attribution, and it introduces a matching evaluation protocol that measures localization fidelity directly.

Across NQ and MS-MARCO, five poisoning attack families, six target LLMs, and multiple traceback baselines, RAGCharacter achieved the best overall trade-off within our benchmark between accurate localization and low over-attribution. These results suggest that prompt-conditioned, black-box character-level traceback can be feasible when attribution is restricted to prompt-used evidence. Our results support a forensic view of poisoned RAG: the primary problem is to attribute a concrete misgeneration event to a precise poisoned span, while remediation is a downstream operational decision that may depend on system-specific utility constraints. More broadly, our findings suggest that practical RAG security should not stop at identifying suspicious documents; it may benefit from isolating small implicated spans that can be masked, audited, or considered for downstream remediation with reduced collateral damage.

\section{Limitations and Future Studies}
\label{sec6}

Our study has several limitations that also define the next steps for character-level RAG forensics.

First, the current Pass-1 procedure is event-conditioned and partly answer-anchored. In practice, the observed wrong answer is still useful for pruning chunks, ordering candidate regions, and constructing cheap attribution shortcuts. This design is effective and cost-efficient, but it may favor poisoning attacks whose injected evidence is lexically explicit. Future work should further weaken this dependence and move toward more counterfactual-first traceback, where candidate selection is driven primarily by event suppression rather than answer matching.

Second, our current formulation mainly assumes that responsibility can be localized to one or a few contiguous spans inside the prompt-used evidence. This is a practical starting point for auditing and remediation, but real poisoning may be distributed across multiple non-adjacent phrases, multiple chunks, or interactions between poisoned and benign evidence. Extending traceback from single-span attribution to structured multi-span and multi-chunk responsibility remains an important direction.

Third, Pass-1 operates on the fixed Pass-0 retrieval trace. This makes the forensic problem reproducible and well-defined, since we analyze only the evidence that actually reached the generator. However, it also means that our results should be interpreted as prompt-conditioned attribution rather than a full guarantee under re-retrieval, re-indexing, or changed chunk boundaries. Future work should evaluate whether span-level sanitization remains effective when the retrieval pipeline is rerun after remediation.

Finally, our empirical study is still limited in breadth. We evaluate two QA corpora, five poisoning attack families, and six target LLMs, which is sufficient to establish the setting but does not exhaust the space of realistic RAG deployments. In addition, the character-level versions of prior traceback baselines are best-effort black-box adaptations rather than canonical span-level methods. Future work should expand the benchmark to more datasets, retrievers, attacks, and adaptive adversaries, while also developing stronger span-level baselines and more efficient replay-based search procedures.

\bibliography{references}

\newpage


\appendix

\section{Character-level Traceback performance data}
\label{trace_back_tables_all}

\begin{table*}[hbpt]
\centering
\caption{Experimental results on the NQ dataset (Gemma).}
\label{tab:nq_gemma}
\renewcommand{\arraystretch}{1.05}
\setlength{\tabcolsep}{3pt}
\resizebox{\textwidth}{!}{%
\begin{tabular}{llccccc}
\toprule
Method & Metric & AbvDecoding & CorpusPoison & GASLITE & PoisonedRAG\_black & PoisonedRAG\_white \\
\midrule
\multirow{3}{*}{RAGForensics}
& Char F1  & 0.087 & 0.137 & 0.034 & 0.133 & 0.093 \\
& Char IoU & 0.046 & 0.075 & 0.017 & 0.072 & 0.049 \\
& Char FPR & 0.954 & 0.925 & 0.983 & 0.928 & 0.951 \\
\cmidrule(lr){1-7}
\multirow{3}{*}{RAGOrigin}
& Char F1  & 0.088 & 0.138 & 0.033 & 0.133 & 0.092 \\
& Char IoU & 0.047 & 0.075 & 0.017 & 0.072 & 0.049 \\
& Char FPR & 0.953 & 0.925 & 0.983 & 0.928 & 0.951 \\
\cmidrule(lr){1-7}
\multirow{3}{*}{RobustRAG}
& Char F1  & 0.095 & 0.133 & 0.046 & 0.140 & 0.105 \\
& Char IoU & 0.050 & 0.071 & 0.023 & 0.076 & 0.055 \\
& Char FPR & 0.950 & 0.929 & 0.977 & 0.924 & 0.945 \\
\cmidrule(lr){1-7}
\multirow{3}{*}{PPL}
& Char F1  & 0.087 & 0.142 & 0.031 & 0.127 & 0.095 \\
& Char IoU & 0.046 & 0.078 & 0.016 & 0.069 & 0.050 \\
& Char FPR & 0.954 & 0.922 & 0.984 & 0.931 & 0.950 \\
\cmidrule(lr){1-7}
\multirow{3}{*}{ours}
& Char F1  & 0.903 & 0.979 & 0.856 & 0.939 & 0.964 \\
& Char IoU & 0.895 & 0.971 & 0.846 & 0.931 & 0.954 \\
& Char FPR & 0.086 & 0.000 & 0.127 & 0.049 & 0.012 \\
\bottomrule
\end{tabular}%
}
\end{table*}

\begin{table*}[hbpt]
\centering
\caption{Experimental results on the NQ dataset (Gemma3).}
\label{tab:nq_gemma3}
\renewcommand{\arraystretch}{1.05}
\setlength{\tabcolsep}{3pt}
\resizebox{\textwidth}{!}{%
\begin{tabular}{llccccc}
\toprule
Method & Metric & AbvDecoding & CorpusPoison & GASLITE & PoisonedRAG\_black & PoisonedRAG\_white \\
\midrule
\multirow{3}{*}{RAGForensics}
& Char F1  & 0.085 & 0.134 & 0.035 & 0.135 & 0.093 \\
& Char IoU & 0.045 & 0.073 & 0.018 & 0.074 & 0.049 \\
& Char FPR & 0.955 & 0.927 & 0.982 & 0.926 & 0.951 \\
\cmidrule(lr){1-7}
\multirow{3}{*}{RAGOrigin}
& Char F1  & 0.083 & 0.135 & 0.035 & 0.135 & 0.092 \\
& Char IoU & 0.043 & 0.074 & 0.018 & 0.073 & 0.049 \\
& Char FPR & 0.957 & 0.926 & 0.982 & 0.927 & 0.951 \\
\cmidrule(lr){1-7}
\multirow{3}{*}{RobustRAG}
& Char F1  & 0.095 & 0.133 & 0.046 & 0.140 & 0.105 \\
& Char IoU & 0.050 & 0.071 & 0.023 & 0.076 & 0.055 \\
& Char FPR & 0.950 & 0.929 & 0.977 & 0.924 & 0.945 \\
\cmidrule(lr){1-7}
\multirow{3}{*}{PPL}
& Char F1  & 0.086 & 0.143 & 0.032 & 0.126 & 0.094 \\
& Char IoU & 0.045 & 0.078 & 0.016 & 0.068 & 0.050 \\
& Char FPR & 0.955 & 0.922 & 0.984 & 0.932 & 0.950 \\
\cmidrule(lr){1-7}
\multirow{3}{*}{ours}
& Char F1  & 0.785 & 0.971 & 0.814 & 0.698 & 0.794 \\
& Char IoU & 0.777 & 0.962 & 0.803 & 0.691 & 0.778 \\
& Char FPR & 0.210 & 0.000 & 0.162 & 0.302 & 0.176 \\
\bottomrule
\end{tabular}%
}
\end{table*}

\begin{table*}[hbpt]
\centering
\caption{Experimental results on the NQ dataset (qwen2.5).}
\label{tab:nq_qwen25}
\renewcommand{\arraystretch}{1.05}
\setlength{\tabcolsep}{3pt}
\resizebox{\textwidth}{!}{%
\begin{tabular}{llccccc}
\toprule
 Method & Metric & AbvDecoding & CorpusPoison & GASLITE & PoisonedRAG\_black & PoisonedRAG\_white \\
\midrule
\multirow{3}{*}{RAGForensics}
& Char F1  & 0.087 & 0.137 & 0.034 & 0.143 & 0.095 \\
& Char IoU & 0.046 & 0.075 & 0.017 & 0.078 & 0.051 \\
& Char FPR & 0.954 & 0.925 & 0.983 & 0.922 & 0.949 \\
\cmidrule(lr){1-7}
\multirow{3}{*}{RAGOrigin}
& Char F1  & 0.084 & 0.137 & 0.035 & 0.137 & 0.094 \\
& Char IoU & 0.044 & 0.075 & 0.018 & 0.075 & 0.050 \\
& Char FPR & 0.956 & 0.925 & 0.982 & 0.925 & 0.950 \\
\cmidrule(lr){1-7}
\multirow{3}{*}{RobustRAG}
& Char F1  & 0.095 & 0.133 & 0.046 & 0.140 & 0.105 \\
& Char IoU & 0.050 & 0.071 & 0.023 & 0.076 & 0.055 \\
& Char FPR & 0.950 & 0.929 & 0.977 & 0.924 & 0.945 \\
\cmidrule(lr){1-7}
\multirow{3}{*}{PPL}
& Char F1  & 0.084 & 0.141 & 0.032 & 0.129 & 0.095 \\
& Char IoU & 0.045 & 0.077 & 0.016 & 0.070 & 0.050 \\
& Char FPR & 0.955 & 0.923 & 0.984 & 0.930 & 0.950 \\
\cmidrule(lr){1-7}
\multirow{3}{*}{ours}
& Char F1  & 0.918 & 0.976 & 0.858 & 0.993 & 0.921 \\
& Char IoU & 0.909 & 0.965 & 0.842 & 0.989 & 0.907 \\
& Char FPR & 0.072 & 0.000 & 0.122 & 0.011 & 0.085 \\
\bottomrule
\end{tabular}%
}
\end{table*}

\begin{table*}[hbpt]
\centering
\caption{Experimental results on the NQ dataset (gpt-4o-mini).}
\label{tab:nq_gpt-4o-mini}
\renewcommand{\arraystretch}{1.05}
\setlength{\tabcolsep}{3pt}
\resizebox{\textwidth}{!}{%
\begin{tabular}{llccccc}
\toprule
 Method & Metric & AbvDecoding & CorpusPoison & GASLITE & PoisonedRAG\_black & PoisonedRAG\_white \\
\midrule
\multirow{3}{*}{RAGForensics}
& Char F1  & 0.086 & 0.131 & 0.034 & 0.131 & 0.093 \\
& Char IoU & 0.045 & 0.071 & 0.017 & 0.071 & 0.050 \\
& Char FPR & 0.955 & 0.929 & 0.983 & 0.929 & 0.950 \\
\cmidrule(lr){1-7}
\multirow{3}{*}{RAGOrigin}
& Char F1  & 0.084 & 0.132 & 0.034 & 0.130 & 0.091 \\
& Char IoU & 0.044 & 0.072 & 0.017 & 0.071 & 0.049 \\
& Char FPR & 0.956 & 0.928 & 0.983 & 0.929 & 0.951 \\
\cmidrule(lr){1-7}
\multirow{3}{*}{RobustRAG}
& Char F1  & 0.094 & 0.156 & 0.042 & 0.136 & 0.091 \\
& Char IoU & 0.050 & 0.085 & 0.021 & 0.074 & 0.048 \\
& Char FPR & 0.950 & 0.915 & 0.979 & 0.926 & 0.952 \\
\cmidrule(lr){1-7}
\multirow{3}{*}{PPL}
& Char F1  & 0.085 & 0.141 & 0.030 & 0.126 & 0.094 \\
& Char IoU & 0.045 & 0.077 & 0.015 & 0.068 & 0.050 \\
& Char FPR & 0.955 & 0.923 & 0.985 & 0.932 & 0.950 \\
\cmidrule(lr){1-7}
\multirow{3}{*}{ours}
& Char F1  & 0.773 & 0.970 & 0.796 & 0.764 & 0.790 \\
& Char IoU & 0.773 & 0.960 & 0.778 & 0.747 & 0.765 \\
& Char FPR & 0.227 & 0.014 & 0.192 & 0.233 & 0.170 \\
\bottomrule
\end{tabular}%
}
\end{table*}

\begin{table*}[hbpt]
\centering
\caption{Experimental results on the NQ dataset (llama3).}
\label{tab:nq_llama3}
\renewcommand{\arraystretch}{1.05}
\setlength{\tabcolsep}{3pt}
\resizebox{\textwidth}{!}{%
\begin{tabular}{llccccc}
\toprule
 Method & Metric & AbvDecoding & CorpusPoison & GASLITE & PoisonedRAG\_black & PoisonedRAG\_white \\
\midrule
\multirow{3}{*}{RAGForensics}
& Char F1  & 0.087 & 0.139 & 0.032 & 0.133 & 0.095 \\
& Char IoU & 0.046 & 0.076 & 0.016 & 0.072 & 0.051 \\
& Char FPR & 0.954 & 0.924 & 0.984 & 0.928 & 0.949 \\
\cmidrule(lr){1-7}
\multirow{3}{*}{RAGOrigin}
& Char F1  & 0.090 & 0.138 & 0.033 & 0.133 & 0.096 \\
& Char IoU & 0.047 & 0.075 & 0.017 & 0.073 & 0.051 \\
& Char FPR & 0.953 & 0.925 & 0.983 & 0.927 & 0.949 \\
\cmidrule(lr){1-7}
\multirow{3}{*}{RobustRAG}
& Char F1  & 0.095 & 0.133 & 0.046 & 0.140 & 0.105 \\
& Char IoU & 0.050 & 0.071 & 0.023 & 0.076 & 0.055 \\
& Char FPR & 0.950 & 0.929 & 0.977 & 0.924 & 0.945 \\
\cmidrule(lr){1-7}
\multirow{3}{*}{PPL}
& Char F1  & 0.086 & 0.144 & 0.031 & 0.126 & 0.095 \\
& Char IoU & 0.045 & 0.079 & 0.016 & 0.068 & 0.050 \\
& Char FPR & 0.955 & 0.921 & 0.984 & 0.932 & 0.950 \\
\cmidrule(lr){1-7}
\multirow{3}{*}{ours}
& Char F1  & 0.886 & 0.973 & 0.878 & 0.929 & 0.961 \\
& Char IoU & 0.879 & 0.963 & 0.867 & 0.920 & 0.952 \\
& Char FPR & 0.104 & 0.001 & 0.107 & 0.064 & 0.027 \\
\bottomrule
\end{tabular}%
}
\end{table*}

\begin{table*}[hbpt]
\centering
\caption{Experimental results on the NQ dataset (gpt-5-mini).}
\label{tab:nq_gpt-5-mini}
\renewcommand{\arraystretch}{1.05}
\setlength{\tabcolsep}{3pt}
\resizebox{\textwidth}{!}{%
\begin{tabular}{llccccc}
\toprule
 Method & Metric & AbvDecoding & CorpusPoison & GASLITE & PoisonedRAG\_black & PoisonedRAG\_white \\
\midrule
\multirow{3}{*}{RAGForensics}
& Char F1  & 0.082 & 0.137 & 0.036 & 0.123 & 0.098 \\
& Char IoU & 0.043 & 0.075 & 0.018 & 0.066 & 0.052 \\
& Char FPR & 0.957 & 0.925 & 0.982 & 0.934 & 0.948 \\
\cmidrule(lr){1-7}
\multirow{3}{*}{RAGOrigin}
& Char F1  & 0.081 & 0.133 & 0.035 & 0.123 & 0.103 \\
& Char IoU & 0.042 & 0.073 & 0.018 & 0.066 & 0.055 \\
& Char FPR & 0.958 & 0.927 & 0.982 & 0.934 & 0.945 \\
\cmidrule(lr){1-7}
\multirow{3}{*}{RobustRAG}
& Char F1  & 0.094 & 0.156 & 0.042 & 0.136 & 0.091 \\
& Char IoU & 0.050 & 0.085 & 0.021 & 0.074 & 0.048 \\
& Char FPR & 0.950 & 0.915 & 0.979 & 0.926 & 0.952 \\
\cmidrule(lr){1-7}
\multirow{3}{*}{PPL}
& Char F1  & 0.088 & 0.141 & 0.038 & 0.128 & 0.112 \\
& Char IoU & 0.046 & 0.077 & 0.019 & 0.069 & 0.060 \\
& Char FPR & 0.954 & 0.923 & 0.981 & 0.931 & 0.940 \\
\cmidrule(lr){1-7}
\multirow{3}{*}{ours}
& Char F1  & 0.982 & 1.000 & 0.955 & 0.978 & 0.985 \\
& Char IoU & 0.974 & 1.000 & 0.925 & 0.967 & 0.973 \\
& Char FPR & 0.000 & 0.000 & 0.075 & 0.015 & 0.027 \\
\bottomrule
\end{tabular}%
}
\end{table*}

\begin{table*}[hbpt]
\centering
\caption{Experimental results on the MS-MARCO dataset (Gemma).}
\label{tab:msmarco_gemma}
\renewcommand{\arraystretch}{1.05}
\setlength{\tabcolsep}{3pt}

\resizebox{\textwidth}{!}{%
\begin{tabular}{llccccc}
\toprule
 Method & Metric & AbvDecoding & CorpusPoison & GASLITE & PoisonedRAG\_black & PoisonedRAG\_white \\
\midrule
\multirow{3}{*}{RAGForensics}
& Char F1  & 0.208 & 0.244 & 0.086 & 0.332 & 0.249 \\
& Char IoU & 0.127 & 0.150 & 0.047 & 0.233 & 0.160 \\
& Char FPR & 0.873 & 0.850 & 0.953 & 0.767 & 0.840 \\
\cmidrule(lr){1-7}
\multirow{3}{*}{RAGOrigin}
& Char F1  & 0.210 & 0.255 & 0.087 & 0.331 & 0.244 \\
& Char IoU & 0.130 & 0.157 & 0.047 & 0.231 & 0.156 \\
& Char FPR & 0.870 & 0.843 & 0.953 & 0.769 & 0.844 \\
\cmidrule(lr){1-7}
\multirow{3}{*}{RobustRAG}
& Char F1  & 0.077 & 0.120 & 0.059 & 0.112 & 0.109 \\
& Char IoU & 0.040 & 0.064 & 0.031 & 0.060 & 0.058 \\
& Char FPR & 0.960 & 0.936 & 0.969 & 0.940 & 0.942 \\
\cmidrule(lr){1-7}
\multirow{3}{*}{PPL}
& Char F1  & 0.180 & 0.268 & 0.074 & 0.277 & 0.225 \\
& Char IoU & 0.108 & 0.168 & 0.040 & 0.184 & 0.141 \\
& Char FPR & 0.892 & 0.832 & 0.960 & 0.816 & 0.859 \\
\cmidrule(lr){1-7}
\multirow{3}{*}{ours}
& Char F1  & 0.855 & 0.969 & 0.713 & 0.894 & 0.829 \\
& Char IoU & 0.833 & 0.959 & 0.687 & 0.888 & 0.802 \\
& Char FPR & 0.097 & 0.000 & 0.207 & 0.073 & 0.086 \\
\bottomrule
\end{tabular}%
}
\end{table*}

\begin{table*}[hbpt]
\centering
\caption{Experimental results on the MS-MARCO dataset (Gemma3).}
\label{tab:msmarco_gemma3}
\renewcommand{\arraystretch}{1.05}
\setlength{\tabcolsep}{3pt}
\resizebox{\textwidth}{!}{%
\begin{tabular}{llccccc}
\toprule
 Method & Metric & AbvDecoding & CorpusPoison & GASLITE & PoisonedRAG\_black & PoisonedRAG\_white \\
\midrule
\multirow{3}{*}{RAGForensics}
& Char F1  & 0.194 & 0.284 & 0.082 & 0.330 & 0.210 \\
& Char IoU & 0.120 & 0.178 & 0.045 & 0.238 & 0.130 \\
& Char FPR & 0.880 & 0.822 & 0.955 & 0.762 & 0.870 \\
\cmidrule(lr){1-7}
\multirow{3}{*}{RAGOrigin}
& Char F1  & 0.200 & 0.276 & 0.083 & 0.332 & 0.215 \\
& Char IoU & 0.124 & 0.172 & 0.045 & 0.238 & 0.133 \\
& Char FPR & 0.876 & 0.828 & 0.955 & 0.762 & 0.867 \\
\cmidrule(lr){1-7}
\multirow{3}{*}{RobustRAG}
& Char F1  & 0.077 & 0.120 & 0.059 & 0.112 & 0.109 \\
& Char IoU & 0.040 & 0.064 & 0.031 & 0.060 & 0.058 \\
& Char FPR & 0.960 & 0.936 & 0.969 & 0.940 & 0.942 \\
\cmidrule(lr){1-7}
\multirow{3}{*}{PPL}
& Char F1  & 0.155 & 0.270 & 0.066 & 0.274 & 0.187 \\
& Char IoU & 0.091 & 0.169 & 0.035 & 0.185 & 0.115 \\
& Char FPR & 0.909 & 0.831 & 0.965 & 0.815 & 0.885 \\
\cmidrule(lr){1-7}
\multirow{3}{*}{ours}
& Char F1  & 0.673 & 0.959 & 0.806 & 0.704 & 0.784 \\
& Char IoU & 0.638 & 0.944 & 0.778 & 0.692 & 0.749 \\
& Char FPR & 0.268 & 0.000 & 0.110 & 0.270 & 0.126 \\
\bottomrule
\end{tabular}%
}
\end{table*}

\begin{table*}[hbpt]
\centering
\caption{Experimental results on the MS-MARCO dataset (qwen2.5).}
\label{tab:msmarco_qwen25}
\renewcommand{\arraystretch}{1.05}
\setlength{\tabcolsep}{3pt}
\resizebox{\textwidth}{!}{%
\begin{tabular}{llccccc}
\toprule
 Method & Metric & AbvDecoding & CorpusPoison & GASLITE & PoisonedRAG\_black & PoisonedRAG\_white \\
\midrule
\multirow{3}{*}{RAGForensics}
& Char F1  & 0.202 & 0.269 & 0.094 & 0.373 & 0.259 \\
& Char IoU & 0.122 & 0.168 & 0.052 & 0.265 & 0.165 \\
& Char FPR & 0.878 & 0.832 & 0.948 & 0.735 & 0.835 \\
\cmidrule(lr){1-7}
\multirow{3}{*}{RAGOrigin}
& Char F1  & 0.196 & 0.266 & 0.091 & 0.343 & 0.228 \\
& Char IoU & 0.121 & 0.164 & 0.050 & 0.246 & 0.142 \\
& Char FPR & 0.879 & 0.836 & 0.950 & 0.754 & 0.858 \\
\cmidrule(lr){1-7}
\multirow{3}{*}{RobustRAG}
& Char F1  & 0.077 & 0.120 & 0.059 & 0.112 & 0.109 \\
& Char IoU & 0.040 & 0.064 & 0.031 & 0.060 & 0.058 \\
& Char FPR & 0.960 & 0.936 & 0.969 & 0.940 & 0.942 \\
\cmidrule(lr){1-7}
\multirow{3}{*}{PPL}
& Char F1  & 0.157 & 0.271 & 0.073 & 0.282 & 0.210 \\
& Char IoU & 0.090 & 0.168 & 0.039 & 0.191 & 0.130 \\
& Char FPR & 0.910 & 0.832 & 0.961 & 0.809 & 0.870 \\
\cmidrule(lr){1-7}
\multirow{3}{*}{ours}
& Char F1  & 0.740 & 0.919 & 0.834 & 0.863 & 0.754 \\
& Char IoU & 0.711 & 0.894 & 0.815 & 0.860 & 0.721 \\
& Char FPR & 0.202 & 0.000 & 0.118 & 0.140 & 0.098 \\
\bottomrule
\end{tabular}%
}
\end{table*}

\begin{table*}[hbpt]
\centering
\caption{Experimental results on the MS-MARCO dataset (gpt-4o-mini).}
\label{tab:msmarco_gpt-4o-mini}
\renewcommand{\arraystretch}{1.05}
\setlength{\tabcolsep}{3pt}
\resizebox{\textwidth}{!}{%
\begin{tabular}{llccccc}
\toprule
 Method & Metric & AbvDecoding & CorpusPoison & GASLITE & PoisonedRAG\_black & PoisonedRAG\_white \\
\midrule
\multirow{3}{*}{RAGForensics}
& Char F1  & 0.155 & 0.222 & 0.097 & 0.346 & 0.199 \\
& Char IoU & 0.091 & 0.135 & 0.054 & 0.246 & 0.122 \\
& Char FPR & 0.909 & 0.865 & 0.946 & 0.754 & 0.878 \\
\cmidrule(lr){1-7}
\multirow{3}{*}{RAGOrigin}
& Char F1  & 0.152 & 0.225 & 0.099 & 0.341 & 0.200 \\
& Char IoU & 0.089 & 0.136 & 0.055 & 0.242 & 0.123 \\
& Char FPR & 0.911 & 0.864 & 0.945 & 0.758 & 0.877 \\
\cmidrule(lr){1-7}
\multirow{3}{*}{RobustRAG}
& Char F1  & 0.095 & 0.354 & 0.048 & 0.144 & 0.099 \\
& Char IoU & 0.050 & 0.223 & 0.025 & 0.079 & 0.053 \\
& Char FPR & 0.950 & 0.777 & 0.975 & 0.921 & 0.947 \\
\cmidrule(lr){1-7}
\multirow{3}{*}{PPL}
& Char F1  & 0.137 & 0.219 & 0.076 & 0.244 & 0.174 \\
& Char IoU & 0.078 & 0.134 & 0.041 & 0.154 & 0.106 \\
& Char FPR & 0.922 & 0.866 & 0.959 & 0.846 & 0.894 \\
\cmidrule(lr){1-7}
\multirow{3}{*}{ours}
& Char F1  & 0.551 & 0.956 & 0.704 & 0.582 & 0.598 \\
& Char IoU & 0.528 & 0.941 & 0.655 & 0.555 & 0.551 \\
& Char FPR & 0.371 & 0.000 & 0.126 & 0.376 & 0.183 \\
\bottomrule
\end{tabular}%
}
\end{table*}

\begin{table*}[hbpt]
\centering
\caption{Experimental results on the MS-MARCO dataset (llama3).}
\label{tab:msmarco_llama3}
\renewcommand{\arraystretch}{1.05}
\setlength{\tabcolsep}{3pt}
\resizebox{\textwidth}{!}{%
\begin{tabular}{llccccc}
\toprule
 Method & Metric & AbvDecoding & CorpusPoison & GASLITE & PoisonedRAG\_black & PoisonedRAG\_white \\
\midrule
\multirow{3}{*}{RAGForensics}
& Char F1  & 0.208 & 0.258 & 0.087 & 0.340 & 0.221 \\
& Char IoU & 0.126 & 0.156 & 0.047 & 0.238 & 0.137 \\
& Char FPR & 0.874 & 0.844 & 0.953 & 0.762 & 0.863 \\
\cmidrule(lr){1-7}
\multirow{3}{*}{RAGOrigin}
& Char F1  & 0.211 & 0.280 & 0.096 & 0.346 & 0.231 \\
& Char IoU & 0.129 & 0.173 & 0.052 & 0.243 & 0.143 \\
& Char FPR & 0.871 & 0.827 & 0.948 & 0.757 & 0.857 \\
\cmidrule(lr){1-7}
\multirow{3}{*}{RobustRAG}
& Char F1  & 0.077 & 0.120 & 0.059 & 0.112 & 0.109 \\
& Char IoU & 0.040 & 0.064 & 0.031 & 0.060 & 0.058 \\
& Char FPR & 0.960 & 0.936 & 0.969 & 0.940 & 0.942 \\
\cmidrule(lr){1-7}
\multirow{3}{*}{PPL}
& Char F1  & 0.184 & 0.241 & 0.071 & 0.280 & 0.212 \\
& Char IoU & 0.109 & 0.146 & 0.038 & 0.184 & 0.132 \\
& Char FPR & 0.891 & 0.854 & 0.962 & 0.816 & 0.868 \\
\cmidrule(lr){1-7}
\multirow{3}{*}{ours}
& Char F1  & 0.804 & 0.950 & 0.786 & 0.923 & 0.831 \\
& Char IoU & 0.780 & 0.934 & 0.757 & 0.916 & 0.798 \\
& Char FPR & 0.157 & 0.000 & 0.137 & 0.056 & 0.086 \\
\bottomrule
\end{tabular}%
}
\end{table*}

\begin{table*}[hbpt]
\centering
\caption{Experimental results on the MS-MARCO dataset (gpt-5-mini).}
\label{tab:msmarco_gpt-5-mini}
\renewcommand{\arraystretch}{1.05}
\setlength{\tabcolsep}{3pt}
\resizebox{\textwidth}{!}{%
\begin{tabular}{llccccc}
\toprule
 Method & metric & AbvDecoding & CorpusPoison & GASLITE & PoisonedRAG\_black & PoisonedRAG\_white \\
\midrule
\multirow{3}{*}{RAGForensics}
& Char F1  & 0.190 & 0.255 & 0.132 & 0.339 & 0.308 \\
& Char IoU & 0.114 & 0.153 & 0.073 & 0.236 & 0.209 \\
& Char FPR & 0.886 & 0.847 & 0.927 & 0.764 & 0.791 \\
\cmidrule(lr){1-7}
\multirow{3}{*}{RAGOrigin}
& Char F1  & 0.185 & 0.257 & 0.129 & 0.338 & 0.306 \\
& Char IoU & 0.111 & 0.155 & 0.071 & 0.236 & 0.208 \\
& Char FPR & 0.889 & 0.845 & 0.929 & 0.764 & 0.792 \\
\cmidrule(lr){1-7}
\multirow{3}{*}{RobustRAG}
& Char F1  & 0.095 & 0.354 & 0.048 & 0.144 & 0.099 \\
& Char IoU & 0.050 & 0.223 & 0.025 & 0.079 & 0.053 \\
& Char FPR & 0.950 & 0.777 & 0.975 & 0.921 & 0.947 \\
\cmidrule(lr){1-7}
\multirow{3}{*}{PPL}
& Char F1  & 0.149 & 0.238 & 0.103 & 0.319 & 0.286 \\
& Char IoU & 0.087 & 0.141 & 0.057 & 0.223 & 0.183 \\
& Char FPR & 0.913 & 0.859 & 0.943 & 0.777 & 0.817 \\
\cmidrule(lr){1-7}
\multirow{3}{*}{ours}
& Char F1  & 0.889 & 0.974 & 0.321 & 0.891 & 0.975 \\
& Char IoU & 0.833 & 0.962 & 0.299 & 0.887 & 0.960 \\
& Char FPR & 0.000 & 0.000 & 0.701 & 0.113 & 0.040 \\

\bottomrule
\end{tabular}%
}
\end{table*}

\begin{table*}[hbpt]
\centering
\caption{Character-level performance comparison of PPL-Character, RAGForensics-Character, and RAGOrigin-Character on the NQ dataset under different attack settings.}
\label{tab:nq_all_methods_char_only}
\scriptsize
\setlength{\tabcolsep}{4pt}
\renewcommand{\arraystretch}{1.05}
\resizebox{\textwidth}{!}{%
\begin{tabular}{lllccccc}
\toprule
Model & Method & Metric & AbvDecoding & CorpusPoison & GASLITE & \shortstack{PoisonedRAG\_black} & \shortstack{PoisonedRAG\_white} \\
\midrule
\multirow{9}{*}{Gemma}
& \multirow{3}{*}{PPL-Character} & Char F1  & 0.356 & 0.500 & 0.265 & 0.420 & 0.347 \\
& & Char IoU & 0.312 & 0.399 & 0.222 & 0.376 & 0.282 \\
& & Char FPR & 0.570 & 0.185 & 0.653 & 0.493 & 0.500 \\
\cmidrule(lr){2-8}
& \multirow{3}{*}{RAGForensics-Character} & Char F1  & 0.487 & 0.693 & 0.553 & 0.351 & 0.596 \\
& & Char IoU & 0.441 & 0.669 & 0.500 & 0.285 & 0.544 \\
& & Char FPR & 0.558 & 0.308 & 0.469 & 0.715 & 0.441 \\
\cmidrule(lr){2-8}
& \multirow{3}{*}{RAGOrigin-Character} & Char F1  & 0.376 & 0.795 & 0.421 & 0.451 & 0.497 \\
& & Char IoU & 0.294 & 0.779 & 0.323 & 0.364 & 0.411 \\
& & Char FPR & 0.705 & 0.192 & 0.663 & 0.636 & 0.589 \\
\midrule

\multirow{9}{*}{Gemma3}
& \multirow{3}{*}{PPL-Character} & Char F1  & 0.380 & 0.490 & 0.315 & 0.504 & 0.405 \\
& & Char IoU & 0.322 & 0.377 & 0.258 & 0.437 & 0.338 \\
& & Char FPR & 0.487 & 0.148 & 0.615 & 0.395 & 0.452 \\
\cmidrule(lr){2-8}
& \multirow{3}{*}{RAGForensics-Character} & Char F1  & 0.507 & 0.733 & 0.514 & 0.354 & 0.509 \\
& & Char IoU & 0.467 & 0.713 & 0.465 & 0.285 & 0.449 \\
& & Char FPR & 0.526 & 0.278 & 0.523 & 0.708 & 0.543 \\
\cmidrule(lr){2-8}
& \multirow{3}{*}{RAGOrigin-Character} & Char F1  & 0.394 & 0.866 & 0.412 & 0.439 & 0.502 \\
& & Char IoU & 0.311 & 0.858 & 0.316 & 0.347 & 0.415 \\
& & Char FPR & 0.689 & 0.142 & 0.665 & 0.653 & 0.585 \\
\midrule

\multirow{9}{*}{qwen2.5}
& \multirow{3}{*}{PPL-Character} & Char F1  & 0.244 & 0.522 & 0.400 & 0.430 & 0.318 \\
& & Char IoU & 0.192 & 0.419 & 0.335 & 0.362 & 0.256 \\
& & Char FPR & 0.652 & 0.111 & 0.487 & 0.385 & 0.517 \\
\cmidrule(lr){2-8}
& \multirow{3}{*}{RAGForensics-Character} & Char F1  & 0.503 & 0.728 & 0.575 & 0.358 & 0.581 \\
& & Char IoU & 0.462 & 0.705 & 0.521 & 0.286 & 0.527 \\
& & Char FPR & 0.537 & 0.287 & 0.448 & 0.714 & 0.456 \\
\cmidrule(lr){2-8}
& \multirow{3}{*}{RAGOrigin-Character} & Char F1  & 0.394 & 0.777 & 0.412 & 0.468 & 0.457 \\
& & Char IoU & 0.305 & 0.762 & 0.311 & 0.378 & 0.373 \\
& & Char FPR & 0.694 & 0.238 & 0.671 & 0.622 & 0.627 \\
\midrule

\multirow{9}{*}{llama3}
& \multirow{3}{*}{PPL-Character} & Char F1  & 0.350 & 0.532 & 0.391 & 0.433 & 0.397 \\
& & Char IoU & 0.292 & 0.425 & 0.319 & 0.360 & 0.312 \\
& & Char FPR & 0.517 & 0.084 & 0.504 & 0.418 & 0.373 \\
\cmidrule(lr){2-8}
& \multirow{3}{*}{RAGForensics-Character} & Char F1  & 0.427 & 0.739 & 0.587 & 0.368 & 0.583 \\
& & Char IoU & 0.384 & 0.717 & 0.538 & 0.300 & 0.526 \\
& & Char FPR & 0.616 & 0.274 & 0.444 & 0.700 & 0.467 \\
\cmidrule(lr){2-8}
& \multirow{3}{*}{RAGOrigin-Character} & Char F1  & 0.352 & 0.850 & 0.402 & 0.452 & 0.530 \\
& & Char IoU & 0.268 & 0.842 & 0.306 & 0.357 & 0.440 \\
& & Char FPR & 0.731 & 0.158 & 0.687 & 0.643 & 0.553 \\
\bottomrule
\end{tabular}
}
\end{table*}

\begin{table*}[hbpt]
\centering
\caption{Character-level performance comparison of PPL-Character, RAGForensics-Character, and RAGOrigin-Character on the MS-MARCO dataset under different attack settings.}
\label{tab:msmarco_all_methods_char_only}
\scriptsize
\setlength{\tabcolsep}{4pt}
\renewcommand{\arraystretch}{1.05}
\resizebox{\textwidth}{!}{%
\begin{tabular}{lllccccc}
\toprule
Model & Method & Metric & AbvDecoding & CorpusPoison & GASLITE & \shortstack{PoisonedRAG\_black} & \shortstack{PoisonedRAG\_white} \\
\midrule
\multirow{9}{*}{Gemma}
& \multirow{3}{*}{PPL-Character} & Char F1  & 0.290 & 0.493 & 0.258 & 0.399 & 0.391 \\
& & Char IoU & 0.240 & 0.389 & 0.200 & 0.345 & 0.328 \\
& & Char FPR & 0.580 & 0.067 & 0.658 & 0.447 & 0.414 \\
\cmidrule(lr){2-8}
& \multirow{3}{*}{RAGForensics-Character} & Char F1  & 0.514 & 0.674 & 0.589 & 0.291 & 0.550 \\
& & Char IoU & 0.473 & 0.628 & 0.523 & 0.207 & 0.486 \\
& & Char FPR & 0.510 & 0.288 & 0.423 & 0.793 & 0.449 \\
\cmidrule(lr){2-8}
& \multirow{3}{*}{RAGOrigin-Character} & Char F1  & 0.494 & 0.383 & 0.439 & 0.427 & 0.376 \\
& & Char IoU & 0.425 & 0.294 & 0.349 & 0.343 & 0.294 \\
& & Char FPR & 0.571 & 0.671 & 0.651 & 0.636 & 0.705 \\
\midrule

\multirow{9}{*}{Gemma3}
& \multirow{3}{*}{PPL-Character} & Char F1  & 0.296 & 0.462 & 0.300 & 0.368 & 0.360 \\
& & Char IoU & 0.236 & 0.375 & 0.252 & 0.308 & 0.304 \\
& & Char FPR & 0.542 & 0.171 & 0.608 & 0.480 & 0.500 \\
\cmidrule(lr){2-8}
& \multirow{3}{*}{RAGForensics-Character} & Char F1  & 0.466 & 0.573 & 0.501 & 0.226 & 0.495 \\
& & Char IoU & 0.423 & 0.522 & 0.436 & 0.154 & 0.417 \\
& & Char FPR & 0.546 & 0.462 & 0.456 & 0.846 & 0.494 \\
\cmidrule(lr){2-8}
& \multirow{3}{*}{RAGOrigin-Character} & Char F1  & 0.421 & 0.647 & 0.304 & 0.419 & 0.397 \\
& & Char IoU & 0.345 & 0.639 & 0.218 & 0.321 & 0.303 \\
& & Char FPR & 0.650 & 0.361 & 0.769 & 0.657 & 0.680 \\
\midrule

\multirow{9}{*}{qwen2.5}
& \multirow{3}{*}{PPL-Character} & Char F1  & 0.289 & 0.427 & 0.201 & 0.352 & 0.299 \\
& & Char IoU & 0.221 & 0.336 & 0.165 & 0.291 & 0.244 \\
& & Char FPR & 0.559 & 0.156 & 0.742 & 0.440 & 0.516 \\
\cmidrule(lr){2-8}
& \multirow{3}{*}{RAGForensics-Character} & Char F1  & 0.400 & 0.661 & 0.494 & 0.302 & 0.487 \\
& & Char IoU & 0.343 & 0.632 & 0.436 & 0.219 & 0.425 \\
& & Char FPR & 0.621 & 0.323 & 0.468 & 0.781 & 0.528 \\
\cmidrule(lr){2-8}
& \multirow{3}{*}{RAGOrigin-Character} & Char F1  & 0.509 & 0.864 & 0.353 & 0.498 & 0.452 \\
& & Char IoU & 0.433 & 0.856 & 0.264 & 0.405 & 0.379 \\
& & Char FPR & 0.545 & 0.144 & 0.717 & 0.595 & 0.574 \\
\midrule

\multirow{9}{*}{llama3}
& \multirow{3}{*}{PPL-Character} & Char F1  & 0.397 & 0.460 & 0.329 & 0.340 & 0.381 \\
& & Char IoU & 0.318 & 0.367 & 0.251 & 0.263 & 0.308 \\
& & Char FPR & 0.454 & 0.074 & 0.555 & 0.494 & 0.482 \\
\cmidrule(lr){2-8}
& \multirow{3}{*}{RAGForensics-Character} & Char F1  & 0.513 & 0.743 & 0.500 & 0.284 & 0.541 \\
& & Char IoU & 0.472 & 0.719 & 0.436 & 0.197 & 0.479 \\
& & Char FPR & 0.497 & 0.230 & 0.480 & 0.803 & 0.431 \\
\cmidrule(lr){2-8}
& \multirow{3}{*}{RAGOrigin-Character} & Char F1  & 0.403 & 0.920 & 0.440 & 0.493 & 0.488 \\
& & Char IoU & 0.328 & 0.911 & 0.356 & 0.402 & 0.400 \\
& & Char FPR & 0.666 & 0.053 & 0.603 & 0.598 & 0.553 \\
\bottomrule
\end{tabular}
}
\end{table*}

\clearpage

\section{Performance data visualization}
\label{app_vis}

\begin{figure}[hbpt]
\centering
\includegraphics[width=\textwidth]{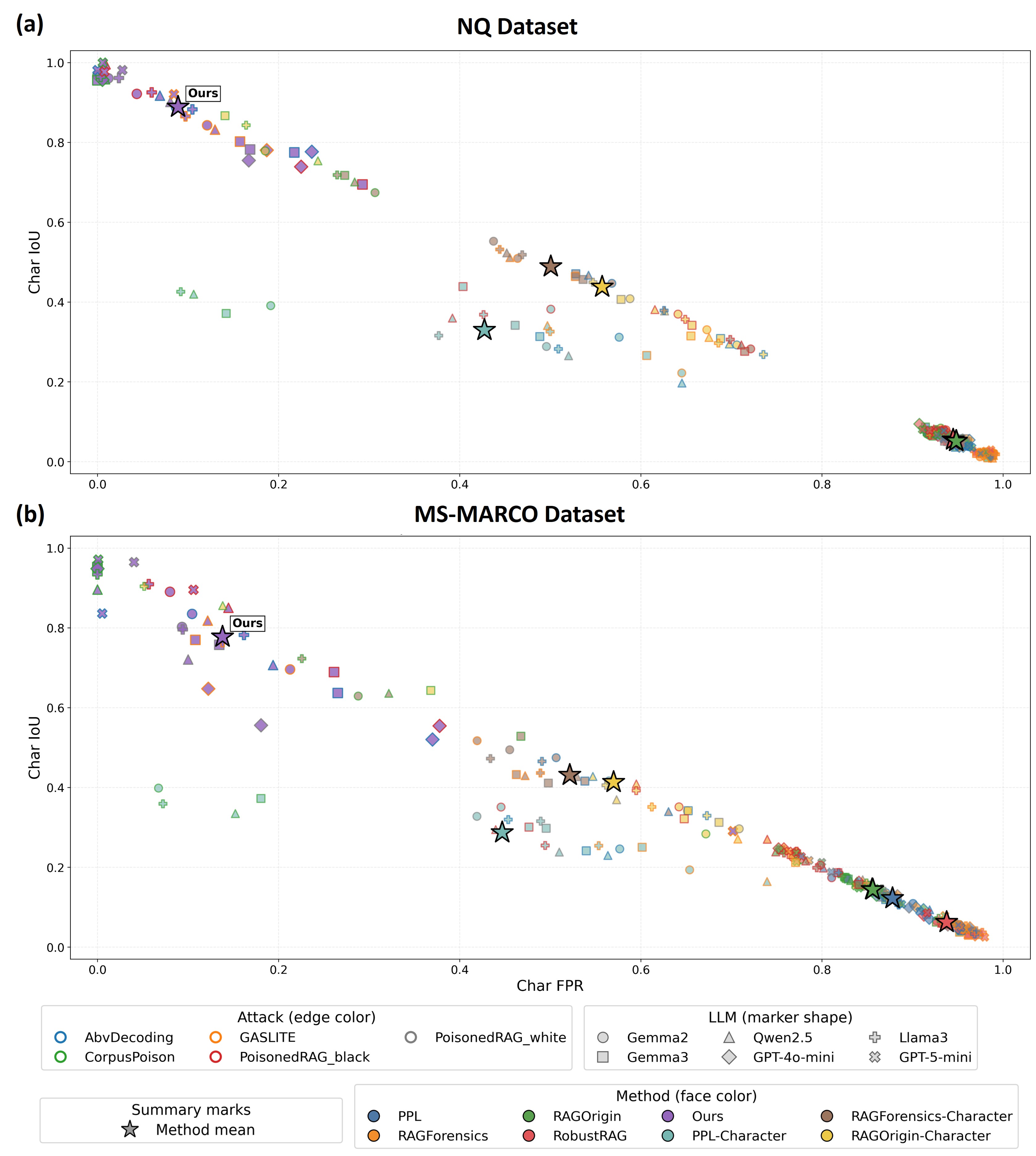}
\caption{Character-level traceback performance across all attacks, target LLMs, and methods, measured by Char IoU and Char FPR. Panel (a) shows NQ and panel (b) shows MS-MARCO. Each raw point corresponds to one attack--LLM--method result. Marker face color denotes method, edge color denotes attack, and marker shape denotes target LLM. Large stars denote method-level averages.}
\label{fig:iou}
\end{figure}

\begin{figure}[hbpt]
\centering
\includegraphics[width=0.8\textwidth]{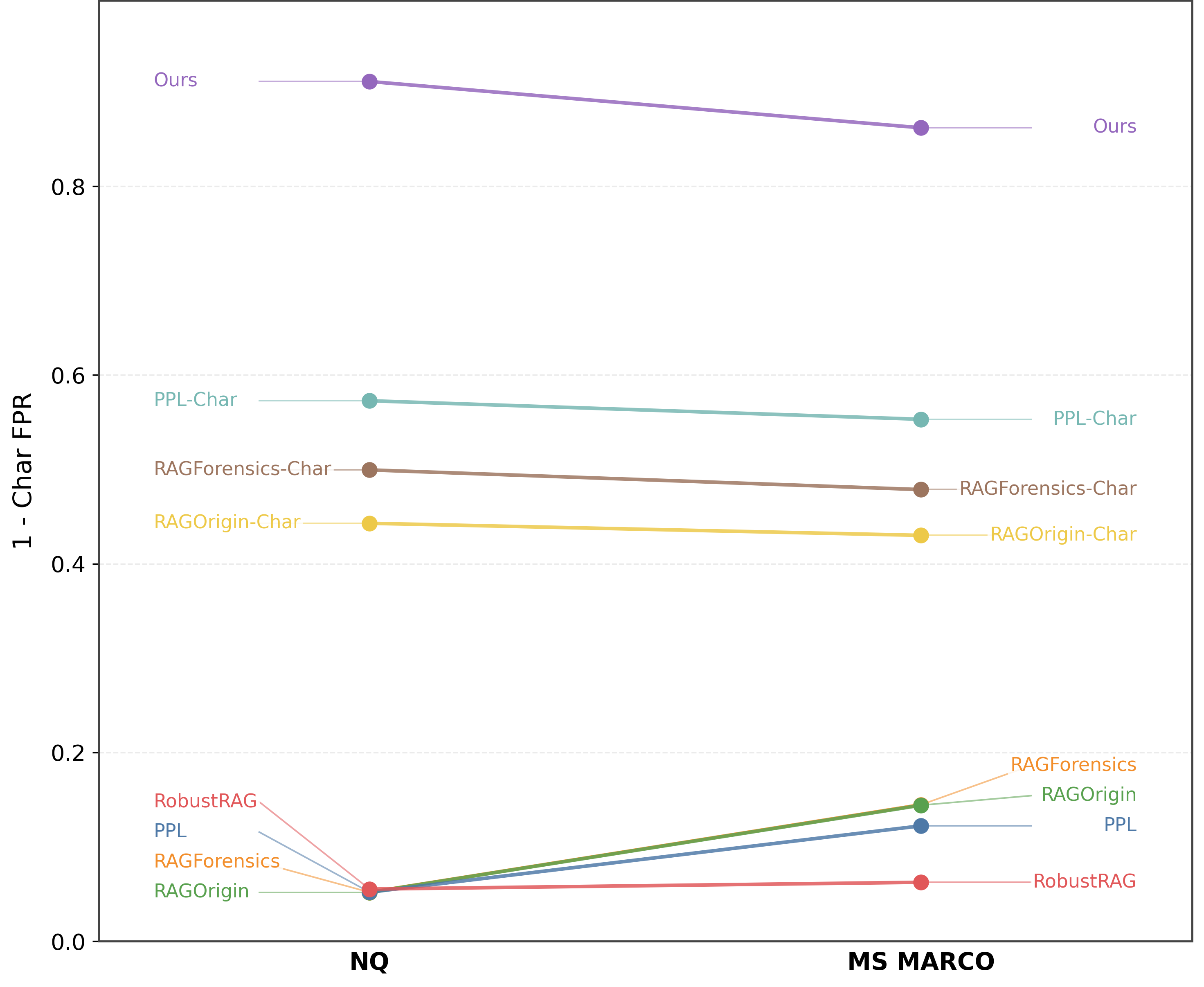}
\caption{Dataset shift by method measured by $1-\text{Char FPR}$, where higher is better and indicates fewer falsely selected characters.}
\label{fig:sg_fpr}
\end{figure}

\begin{figure}[hbpt]
\centering
\includegraphics[width=\textwidth]{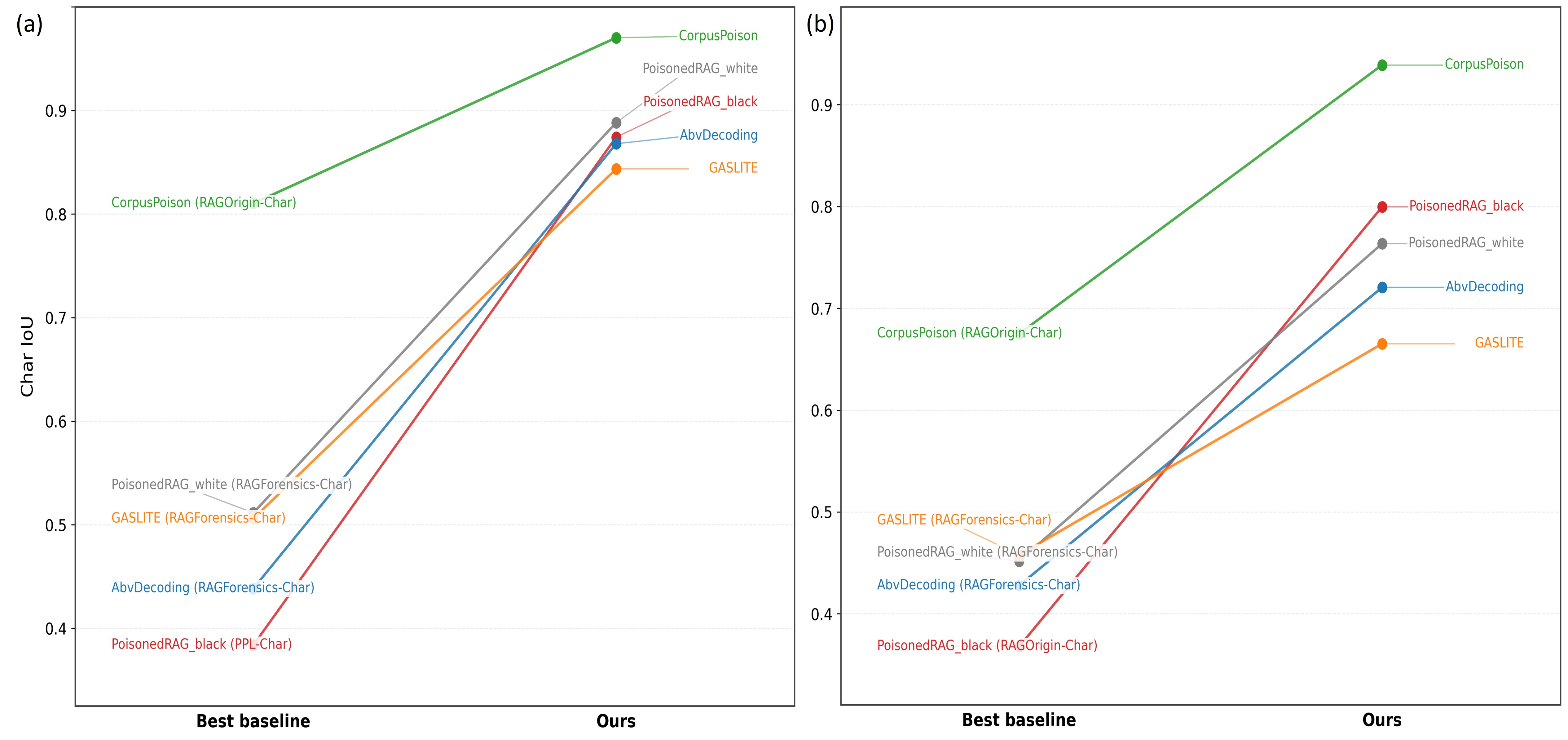}
\caption{Per-attack comparison between our method and the strongest baseline under Char IoU on (a) NQ dataset and (b) MS-MARCO dataset. Each line corresponds to one attack, connecting the best baseline on the left and our method on the right.}
\label{fig:sg_ours}
\end{figure}

\begin{figure}[hbpt]
\centering
\includegraphics[width=\textwidth]{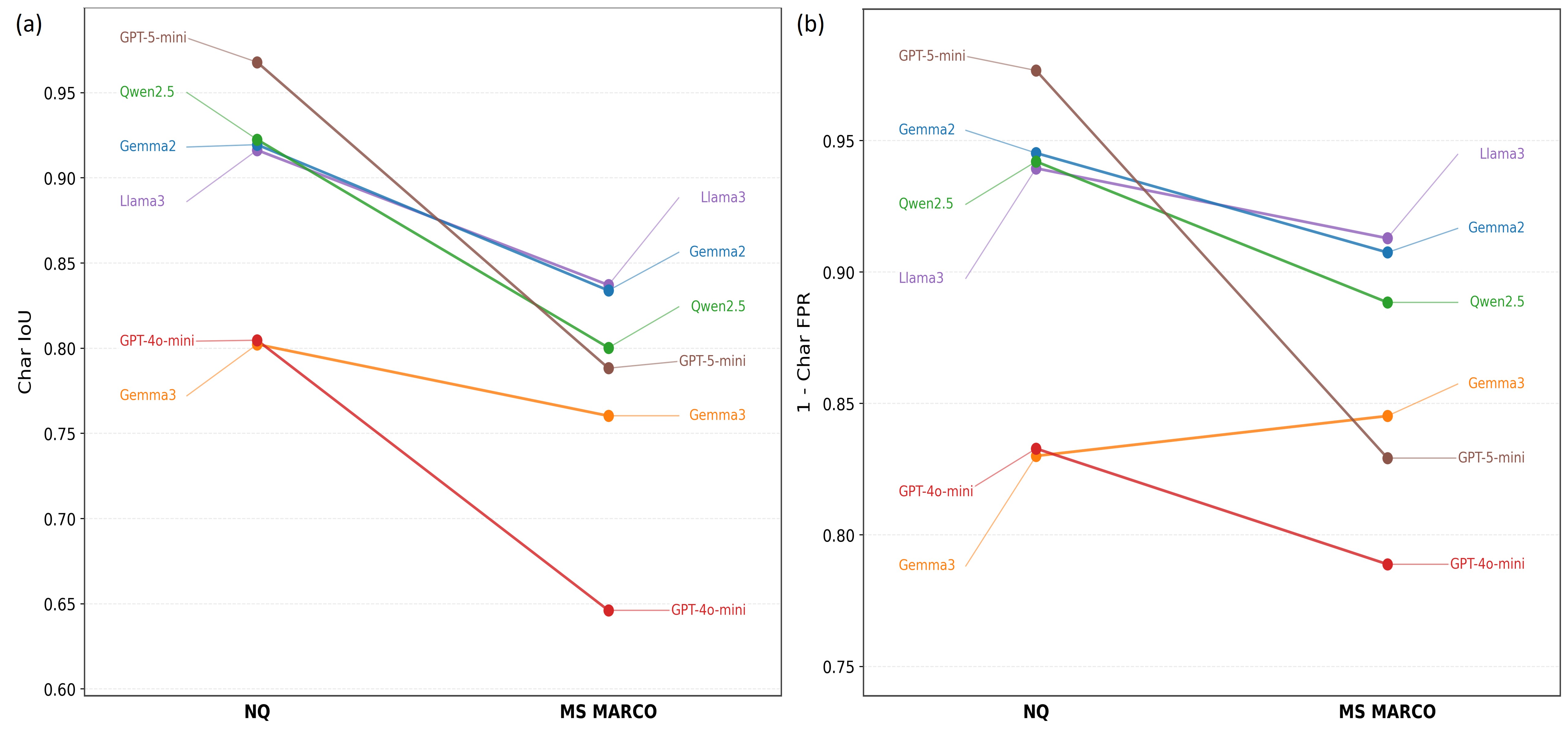}
\caption{Dataset shift across target LLMs for our method under (a) Char IoU and (b) $1-\text{Char\_FPR}$. Each line corresponds to one target LLM, connecting performance from NQ to MS-MARCO.}
\label{fig:sg_ours_llm}
\end{figure}

\clearpage

\end{document}